\documentclass{aa}  

\usepackage{graphicx}	
\usepackage{amsmath}	
\usepackage{txfonts}
\usepackage{latexsym}
\usepackage[dvipsnames]{xcolor}
\usepackage{chngcntr}

\DeclareMathOperator\erf{erf}

\begin{document}

\title{VLT VIMOS Integral Field Spectroscopy of the nova remnant FH\,Ser}

\author{M.~A.~Guerrero
          \inst{1}
          \and
          E.~Santamar\'\i a\inst{2}
          \and
          L.~Takeda\inst{3}
          \and
          J.~I.~Gonz\'{a}lez-Carbajal\inst{4}
          \and
          S.~Cazzoli\inst{1}
          \and 
          A.~Ederoclite\inst{5}
          \and 
          J.~A.~Toal\'{a}\inst{2}
          }

\institute{Instituto de Astrof{\'i}sica de Andaluc{\'i}a, IAA-CSIC, Glorieta de la Astronom\'{i}a S/N, Granada 18008, Spain \\
\email{mar@iaa.es}
\and
Instituto de Radioastronom\'{i}a y Astrof\'{i}sica, Universidad Nacional Aut\'{o}noma de M\'{e}xico, Morelia 58089, Mich., Mexico
\and
Instituto de Astronomia, Geof\'{i}sica e Ci\^{e}ncias Atmosf\'{e}ricas, Universidade de S\~{a}o Paulo, Rua do Mat\~{a}o 1226, S\~{a}o Paulo 05508-900, Brazil
\and
Instituto de Ciencias Nucleares, Universidad Nacional Aut\'{o}noma de M\'{e}xico, Ciudad de M\'{e}xico 04510, M\'{e}xico
\and
Centro de Estudios de F\'{i}sica del Cosmos de Arag\'{o}n, Unidad Asociada al CSIC, Plaza San Juan 1, Teruel 44001, Spain
}

\date{Received September 15, 2024; accepted }

 
\abstract
{ 
FH\,Ser experienced a slow classical nova outburst in February 1970 that was the first one observed at UV, optical, and IR wavelengths. 
Its nova remnant is elliptical in shape, with multiple knots, and a peculiar ring-like filament along its minor axis.
}{ 
This work aims at unveiling the true 3D spatio-kinematical structure of FH\,Ser to investigate the effects of early shaping and to assess its mass and kinetic energy.
}{ 
Very Large Telescope (VLT) VIsible MultiObject Spectrograph (VIMOS) integral field spectroscopic observations 
of FH\,Ser have been obtained.  
The data cube has been analyzed using 3D visualisations that reveal different structural components.  
A simple geometrical model has been compared to the 3D data cube to determine the spatio-kinematic properties of FH\,Ser.
}{
FH\,Ser consists of a tilted prolate ellisoidal shell, most prominent in H$\alpha$, and a ring-like structure, most prominent in [N~{\sc ii}]. 
The ellipsoidal shell has equatorial and polar velocities of $\simeq$505 and $\approx$630 km~s$^{-1}$, respectively, with its major axis tilted by $\simeq52^\circ$ with respect to the line of sight.  
The inclination angle of the symmetry axis of the ring is similar, i.e., it can be described as an equatorial belt of the main ellipsoidal shell. 
FH\,Ser has an ionized mass of $2.6\times10^{-4}$ M$_\odot$, with a kinetic energy of $1.6\times10^{45}$ erg. 
}{
The presence of two different structural components in FH\,Ser with similar orientation can be linked to a density enhancement along a plane, most likely the orbital plane at the time of the nova event.  
The acquisition of integral field spectroscopic observations of nova remnants is most required to disentangle different structural components and to assess their 3D physical structure.   
}

\keywords{Stars: novae, cataclysmic variables -- ISM: individual objects: FH\,Ser}

\maketitle



\section{Introduction}

A classical nova (CN) explosion is the outcome of interactions in a close binary system, where a red dwarf or red giant transfers hydrogen-rich material to a white dwarf (WD) via an accretion disk\footnote{Helium-rich material is also transferred in certain systems \citep[see, e.g.,][]{2003ApJ...598L.107K}.}. 
When the material accreted onto the WD reaches a critical mass, the star undergoes a thermonuclear runaway (TNR) event \citep{Starrfield+2016}.
During the most active convective stage of the TNR, deep mixing enriches the ejecta with matter from the outer layers of the WD, providing additional fuel for enhanced burning \citep{Starrfield+2020}. 
The energy release ejects significant amounts \citep[10$^{-5}$--10$^{-4}$ M$_\odot$;][]{Gehrz1999} of highly processed material \citep[e.g.][]{JoseHernanz1998,Starrfield+2016,Starrfield+2020} at high speeds \citep[from $\sim$100 to $\simeq$1000 km~s$^{-1}$;][]{BodeEvans2012}.

The turbulent nature of the TNR is expected to produce filaments and clumps with observed inhomogeneous abundance patterns in the nova ejecta \citep{Casanova+2018}.
The interactions of the prompt ejecta at the TNR, the prolonged thick stellar wind that follows it, the accretion disk, the binary companion, and the circumstellar material further shape the nova remnant. 
Indeed, multi-wavelength imaging studies of nova remnants have shown that the ejecta can be highly asymmetric \citep{Chesneau+2008,Shore2013}, whereas spectroscopic observations reveal dust grains \citep{Gehrz+1992,Mason+1996,Evans+2005} and uncondensed metallic gas phase elements \citep{Greenhouse+1990,Hayward+1996}. 
These may be caused by actual abundances gradients, but also be an effect of non-isotropic ionization.

High- to intermediate-dispersion integral field spectroscopic (IFS) observations of nova remnants offer the possibility to investigate their 3D physical structure as they provide kinematical information along the line of sight at each nebular position, but these studies are still scarce \citep[see, e.g.,][]{Celedon+2024}. 
The recent VLT MUSE IFS investigation of the nebular remnant associated with the recurrent nova T\,Pyx, for instance, confirms the great potential of this technique \citep{Santamaria2024}. 
The 3D physical structure of the recurrent nova T\,Pyx is there revealed with unprecedented detail, disclosing a toroidal-like structure in H$\beta$ and two open bowl-shaped bipolar lobes in the [O\,{\sc iii}] emission lines.

Ejecta associated with different outburst can also be isolated in these spatially resolved observations \citep{Izzo+2024}. 
Another case study is that of QU\,Vul, where the GTC MEGARA IFS description of the inhomogeneous shell provided the means to assess its mass, free from assumptions on the macroscopic distribution of material within the shell, whereas the spatially-resolved kinematics makes possible to interpret the typical ``castellated'' velocity line profiles observed in young nova remnants \citep{Santamaria+2022}. 
Similarly a near-IR IFS investigation of the nova remnant V5668\,Sgr revealed notorious ionization degree asymmetries, with low ionization lines enhanced at the polar caps and equatorial torus, while high ionization lines were enhanced only at the polar caps \citep{Takeda+2022}.

\begin{figure*}
\begin{center}
\includegraphics[width=1.0\linewidth]{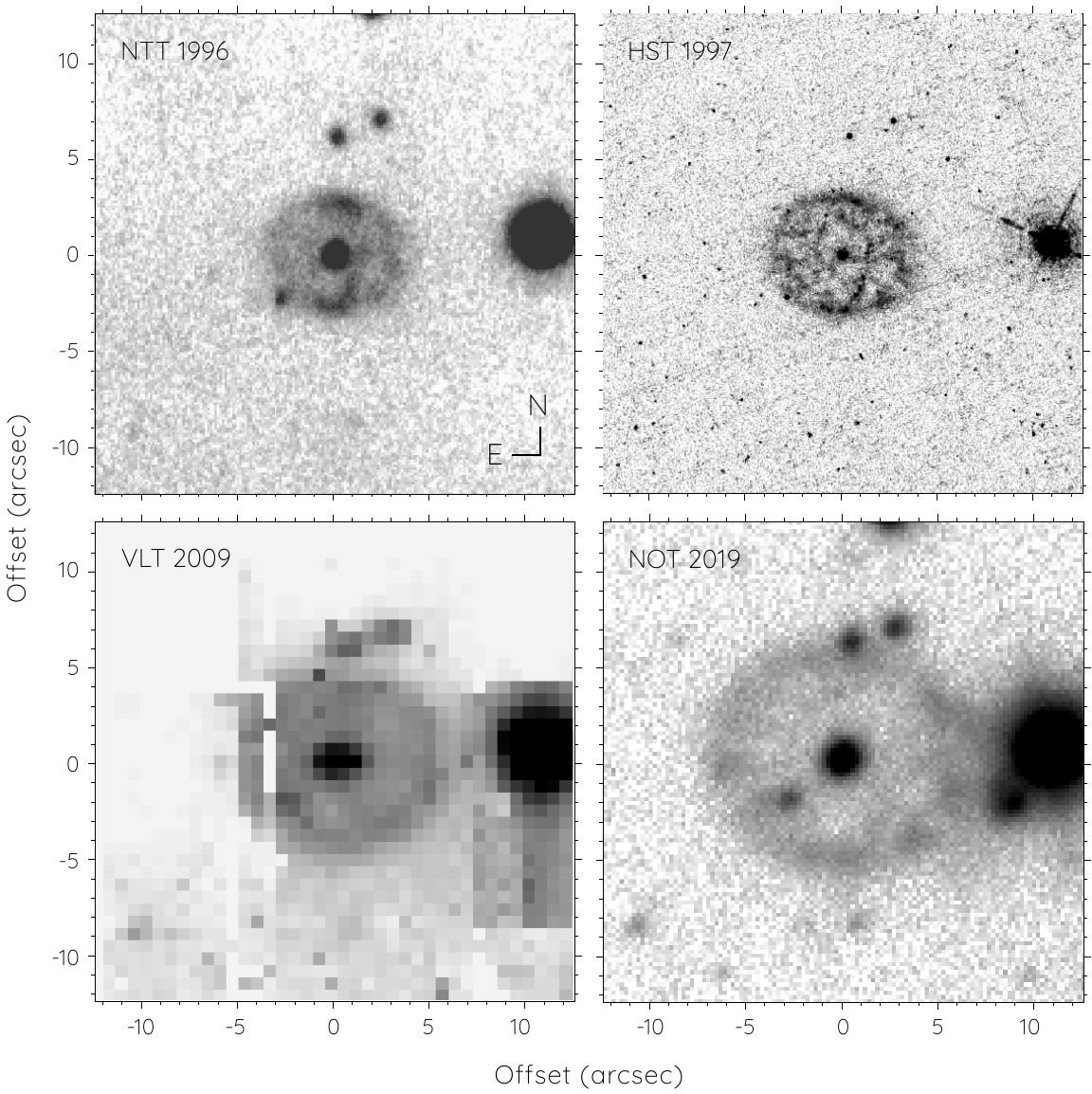}
\caption{
Multi-epoch nebular images of FH Ser. 
The different panels show NTT SUSI H$\alpha$+[N~{\sc ii}] (top-left), HST WFPC2-PC F656N H$\alpha$ (top-right), VLT VIMOS H$\alpha$+[N~{\sc ii}] (bottom-left), and NOT ALFOSC H$\alpha$ (bottom-right)images. 
In all panels north is up and east is to the left. 
The field of view is the same in all panels, with the origin defined to be the position of the central star. 
}
\label{fig:img}
\end{center}
\end{figure*}

In this paper we present an IFS study of the nova remnant FH\,Ser.  
The nova outburst was discovered on 1970 February 13 \citep[$t_0=1970.12$;][]{HiroseHonda1970}, raising from a pre-outburst $m_\mathrm{v}$ magnitude of 16.1 \citep{BurkheadSeeds1970} up to 4.4 mag about five days later by February 18 \citep{BorraAndersen1970}.  
FH\,Ser is considered an archetype of a slow CN \citep{Hack+1993}, with a decline time of three magnitudes from peak ($t_3$) of 62 days \citep{Burkhead+1971,Rosino+1986}. 
The nova entered the nebular phase on 1970 August 13 \citep{Anderson+1971}, i.e.\ about 180 days after its outburst. 
FH\,Ser, being the first nova with optical, IR, and UV coverage \citep{HN1970,GC1974} at the time of its outburst, is considered a case study of nova multi-wavelength research.

Since the nova outburst, the nebular remnant of FH\,Ser has remained in free expansion, with the first resolved H$\alpha$ image of a round shell with an equatorial enhancement or belt obtained on August 1989 \citep{Duerbeck1992} and the last available H$\alpha$ image displaying an ellipsoidal shell obtained on June 2018 \citep{Santamaria+2020}. 
The best quality image of FH\,Ser is undoubtedly that obtained with the Hubble Space Telescope (HST) on 1997 May, using the Wide-Field Planetary Camera 2 (WFPC2) and the F656N H$\alpha$ filter \citep{Gill_OBrian_2000}. 
This image, presented in Fig.~\ref{fig:img}, reveals FH\,Ser as a clumpy limb-brightened ellipse with an equatorial ring, as previously suggested by \citet{Duerbeck1992}.

\citet{Gill_OBrian_2000} assumed this equatorial structure to be a tilted circular ring to estimate an inclination angle $\approx62^\circ$ with the plane of the sky. 
They also presented the only available spatio-kinematic observations and modeling of FH\,Ser based on their HST H$\alpha$ image and William Herschel Telescope (WHT) 
long-slit spectra of the H$\alpha$ and [N~{\sc ii}] $\lambda\lambda$6548,6584 emission lines obtained along the minor and major axes of the elliptical shell.  
Their conjoint model of images and spectra described FH\,Ser as a prolate ellipsoidal shell with a [N~{\sc ii}]-enhanced equatorial ring.  
In this model, the ellipsoidal shell has an axial ratio $1.3\pm0.1$, is tilted by $62^\circ\pm4^\circ$ with the line of sight, and has an equatorial velocity of $490\pm20$ km~s$^{-1}$.

The analysis of the IFS data of FH\,Ser presented in this work helps us peer into the true morpho-kinematics of this nova shell.
This paper is organized as follows. 
In Section~2 we present the details of the IFS data and optical images together with a description of the reduction procedure.  
The data products are presented in Section~3, the results derived from these are described in Section~4 and then discussed in Section~5. 
The main conclusions and a summary of the results are finally presented in Section~6.  
The paper also includes two appendixes that develop two independent techniques to fit the spatio-kinematical information available for FH\,Ser.


\section{Observations and data reduction}

\subsection{Integral field spectroscopic data}

We observed FH\,Ser with the Very Large Telescope (VLT)  VIsible MultiObject Spectrograph (VIMOS) Integral Field Unit (IFU) at the Cerro Paranal European Southern Observatory on 2009 June 27 using the HR Blue (0.571 \AA~pix$^{-1}$), HR Orange~+~GG435 (0.6 \AA~pix$^{-1}$),  and HR Red~+~GG475 (0.6 \AA~pix$^{-1}$) high resolution grisms ($R \sim 3000$).
The spatial sampling is 0.66 arcsec~fiber$^{-1}$. 
We obtained 3 dithered exposures of 1800 s for each grism, and 30 s exposures of the spectro-photometric standard star CD+32\,9927 with the red and orange grisms, and an 84 s exposure of the spectro-photometric standard star Feige~110 with the blue grism.

We used the \texttt{EsoReflex} pipeline \citep{2013AA...559A..96F} to apply basic reduction steps and perform the flux calibration.  
The result was four independent quadrants for each HR grating.  
A series of scripts within the Interactive Data Analysis\footnote{http://www.harrisgeospatial.com/SoftwareTechnology/IDL.aspx (IDL).} (IDL) environment were further used to assemble these four quadrants and to combine the dithered images using the position tables for the HR data.  
The mosaicking of the Southeast quadrant with the adjacent Northeast and Southwest quadrants resulted in a number of faulty pixels along the line and column where they join, which were subsequently assigned null values.

The VLT VIMOS IFS observations of FH\,Ser detect the H$\alpha$ and [N~{\sc ii}] $\lambda\lambda$6548,6584 emission lines both in the HR Orange and HR Red data cubes.  
The H$\alpha$ line is notably brighter than the [N~{\sc ii}] emission lines.  
A visual inspection of the spectra on both data cubes on a spaxel-by-spaxel basis indicates that the emission lines in the HR Orange have generally higher S/N than in the HR Red data cube. 
The spectral analyses hereafter will thus focus on the HR Orange data cube.

\subsection{Direct images}

Direct images of FH\,Ser will be used here for comparison with the VLT VIMOS IFS observations.  
In particular the archival 100~s exposure H$\alpha$+[N~{\sc ii}] image acquired with the 3.58~m New Technology Telescope (NTT) SUper Seeing Instrument (SUSI) at La Silla Observatory on 1996 March 18 \citep{dellaValle1997} and the 2400~s exposure HST WFPC2-PC image acquired with the F656N filter on 1997 May 11 \citep[Prop.\,ID 6770, PI O'Brian;][]{Gill_OBrian_2000}, and our proprietary Nordic Optical Telescope (NOT, Observatory of Roque de los Muchachos) ALhambra Faint Object Spectrograph and Camera (ALFOSC) H$\alpha$ image acquired on 2018 June 6 with an exposure time of 3600 s \citep[see for 
further details in][]{Santamaria+2020}. 
The images are presented in Fig.~\ref{fig:img}.

\section{Data products}

\subsection{H$\alpha$+[N~{\sc ii}] maps}

The spatial distributions of the H$\alpha$+[N~{\sc ii}]
emission lines is compared with available archival images of FH\,Ser in Fig.~\ref{fig:img}.   
The images show that the remnant can be described as an ellipsoidal clumpy shell with the main axis oriented along position angle (PA) $\approx86^\circ$.  
The minor axis of the ring-like structure is basically orthogonal to this direction, with an angular size 5\farcs8$\times$2\farcs5 in the HST image. 
The axial ratio of FH\,Ser derived from the best quality HST WFPC2-PC image \citep[$t=1997.36$, $\Delta t=27.24$ yr;][]{Gill_OBrian_2000} is 6\farcs7$\times$5\farcs8 $\simeq$1.16.

The multi-epoch images presented in Fig.~\ref{fig:img} cover the time-lapse from 1996 to 2019 ($\approx23$ years). 
The angular expansion of the remnant is obvious, but the morphology does not show dramatic changes once the different spatial resolutions of these images are considered. 
\citet{Santamaria+2020} reported angular expansions of (0.125$\pm$0.002)$\times$(0.109$\pm$0.002) arcsec~yr$^{-1}$ for the major and minor axes, implying an axial ratio of $\simeq$1.15, which is consistent with that of the HST WFPC2-PC image.  
The expected angular size of the VLT VIMOS 2009.49 image ($\Delta t=39.37$ yr) would then be $\approx$9\farcs8$\times$8\farcs6, in agreement with its H$\alpha$+[N~{\sc ii}] line map (bottom-left panel of Fig.~\ref{fig:img}). 
%



\subsection{H$\alpha$+[N~{\sc ii}] spectral profile} 

\begin{figure}
\begin{center}
\includegraphics[width=1.0\linewidth]{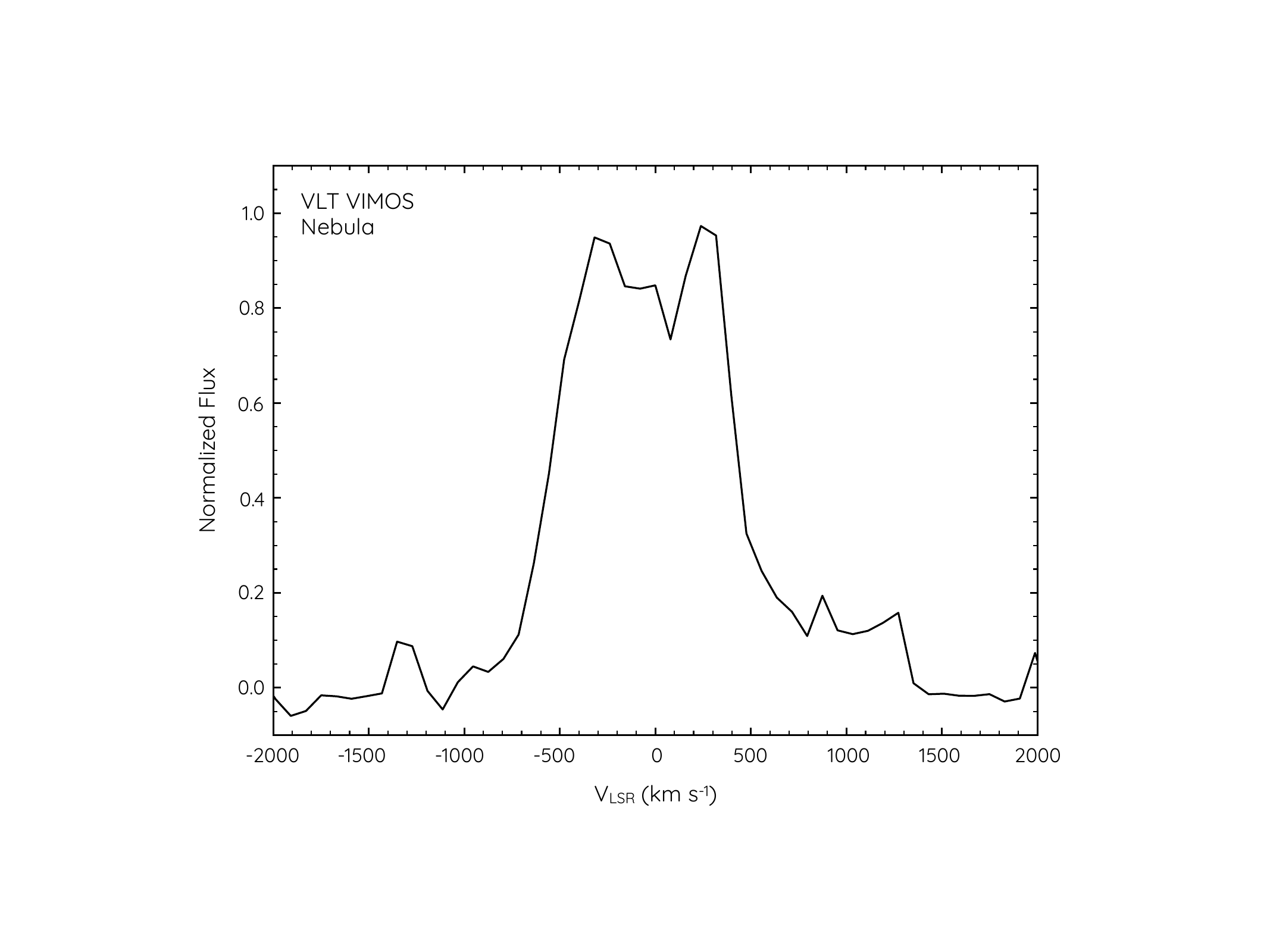}
\caption{
Normalized H$\alpha$+[N~{\sc ii}] $\lambda\lambda$6548,6584 line profile of the integrated nebular emission of FH\,Ser extracted from the VLT VIMOS observations. 
}
\label{fig:1D_spec}
\end{center}
\end{figure}

The integrated H$\alpha$ and [N~{\sc ii}] $\lambda\lambda$6548,6584 line profiles of the nebular component of FH\,Ser are shown in Fig.~\ref{fig:1D_spec}. 
Velocities with respect to the H$\alpha$ line ($\lambda$=6562.849~\AA) are referred to the Local Standard of Rest (LSR). 
The three emission lines overlap notably, with a main double-peak at $\approx\pm300$ km~s$^{-1}$, a faint red wing  between $+800$ and $+1350$ km~s$^{-1}$, and a much fainter blue shoulder\footnote{
The apparent emission line at $\approx-1300$ km~s$^{-1}$ is spurious, actually corresponding to a number of hot pixels projected within the shell of FH\,Ser.  
} 
down to $\approx-1050$ km~s$^{-1}$.

The emission line profile in Fig.~\ref{fig:1D_spec} implies a blue edge for the [N~{\sc ii}] $\lambda$6548 emission line at $-375$ km~s$^{-1}$ and a red edge for the [N~{\sc ii}] $\lambda$6584 emission line at $+405$ km~s$^{-1}$, adopting theoretical wavelengths of $\lambda$=6548.05~\AA\ and $\lambda$=6583.45~\AA\ for the [N~{\sc ii}] $\lambda$6548 and [N~{\sc ii}] $\lambda$6584 emission lines, respectively.   
Under reasonable assumption that both [N~{\sc ii}] emission lines have similar line profiles, then the H$\alpha$ line profile is contaminated at velocities $\leq-270$ km~s$^{-1}$ by the [N~{\sc ii}] $\lambda$6548 emission line and at velocities $\geq+535$ km~s$^{-1}$ by the [N~{\sc ii}] $\lambda$6584 emission line.

\begin{figure*}
\begin{center}
\includegraphics[width=1.0\linewidth]{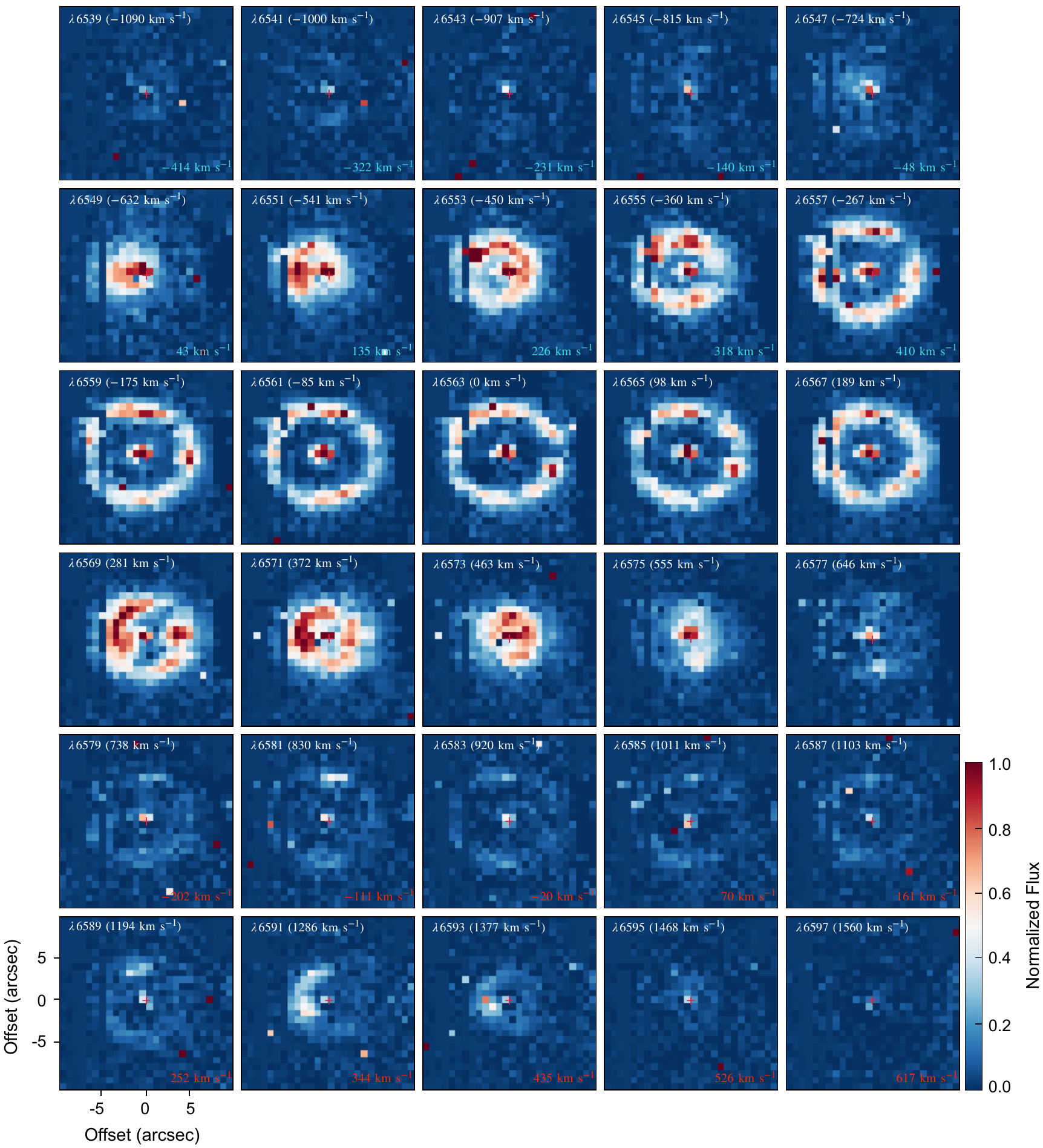}
\caption{
VLT VIMOS tomography of the H$\alpha$ and [N~{\sc ii}] $\lambda\lambda$6548,6584 emission lines of FH\,Ser.  
Each velocity channel has a width of 82 km~s$^{-1}$, i.e., three spectral channels.  
The white, cyan, and red labels correspond to LSR velocities of the H$\alpha$, [N~{\sc ii}] $\lambda$6548 and [N~{\sc ii}] $\lambda$6584 emission lines, respectively. 
The location of the central star is marked by a red ``+'' sign that serves as fiducial point in all channel maps. 
In all maps North is up, East to the left. 
See text for further details on the spectral and spatial extent of each emission line.
}
\label{fig:channel}
\end{center}
\end{figure*}

\begin{figure*}
\begin{center}
\includegraphics[width=1.0\linewidth]{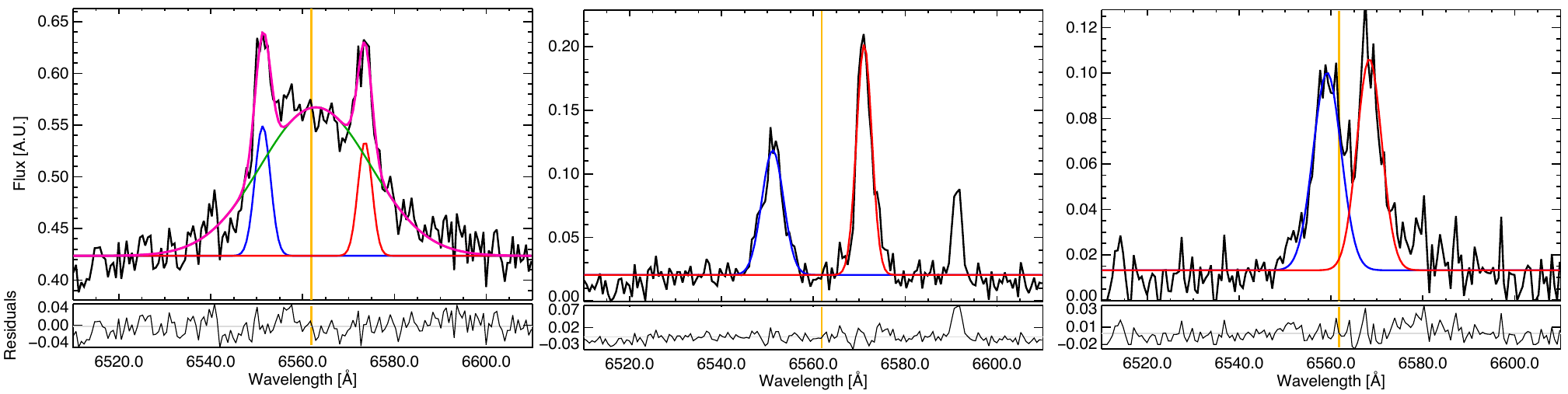}
\caption{
Examples of emission line spectra from the VIMOS cube (black line) and their multi-Gaussian fit (color lines) for the position of the central star (left panel), shell interior (middle panel), and shell edge (right panel). 
The blue and red Gaussian curves correspond to the two nebular components, while the green curve in the left panel represents a broad stellar component. 
Residuals from the fit are shown below each panel. 
As reference, the solid yellow vertical line in all panels marks the systemic wavelength of H$\alpha$. 
}
\label{fig:1Dfit}
\end{center}
\end{figure*}

\begin{figure*}
\begin{center}
\includegraphics[width=1.0\linewidth]{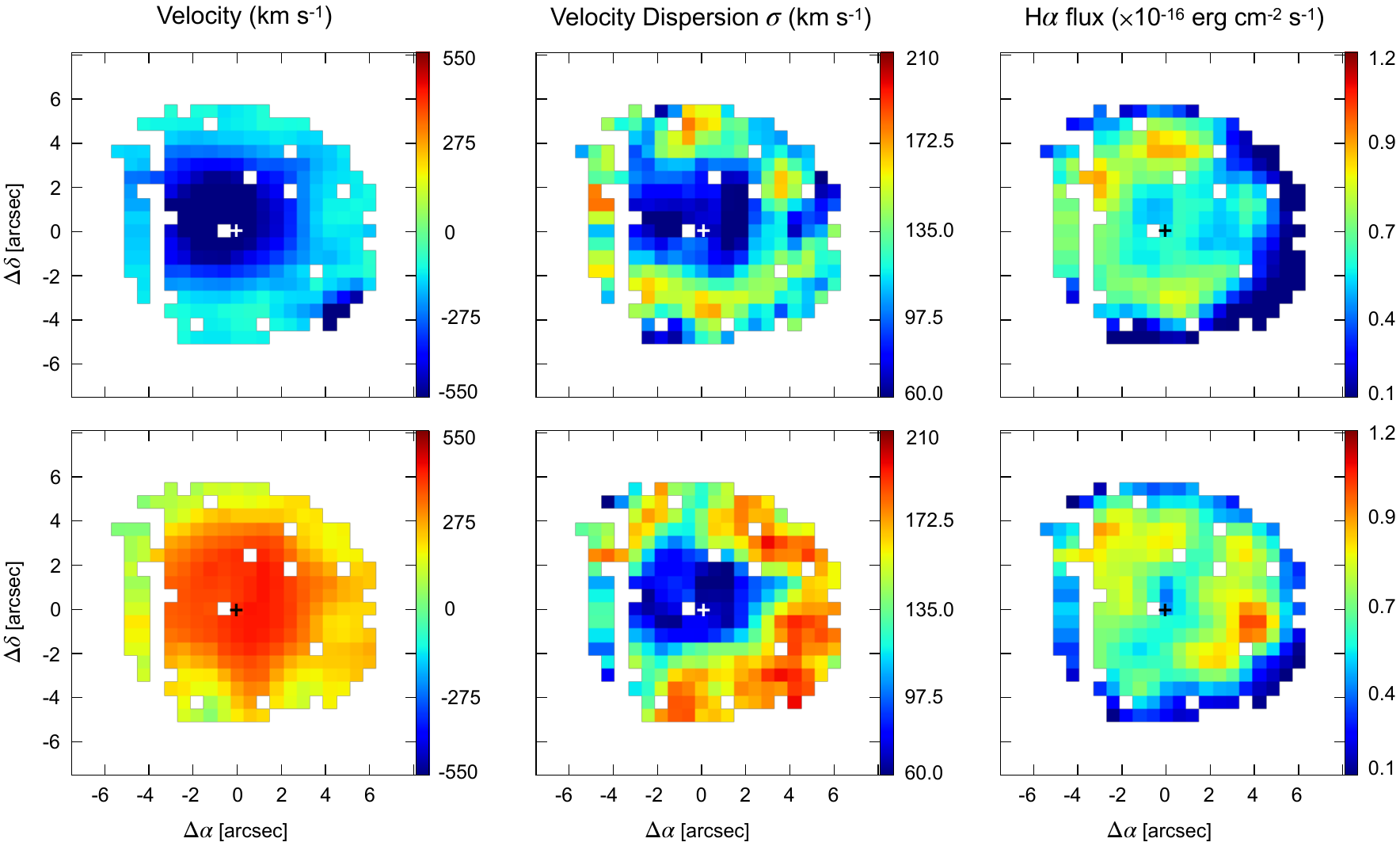}
\caption{
H$\alpha$ velocity (left), velocity dispersion $\sigma$ (middle), and flux intensity (right) 
emission line maps of FH\,Ser obtained from the multi-Gaussian fit of individual spaxels in the VLT VIMOS HR Orange data cube.  
The top and bottom panels show the blue and red nebular components, respectively.  
The cross sign marks the position of the central star (0,0), which is spatially unresolved in the VIMOS data. In all maps, North is up, East to the left.}  
\label{fig:2Dfit}
\end{center}
\end{figure*}

\subsection{H$\alpha$ and [N~{\sc ii}] channel maps}

The H$\alpha$ and [N~{\sc ii}] $\lambda\lambda$6548,6584 line profiles of the nebular components of FH\,Ser in Fig.~\ref{fig:1D_spec} have been used to extract the channel maps shown in Fig.~\ref{fig:channel}. 
Each channel map has a velocity width of 82 km~s$^{-1}$ corresponding to three spectral channels of 0.6 \AA\ at the wavelength of H$\alpha$.  
The central velocity of each channel in the LSR is labeled for the H$\alpha$ (white), [N~{\sc ii}] $\lambda$6548 (cyan), and [N~{\sc ii}] $\lambda$6584 (red) emission lines.

As noted above, the H$\alpha$ and [N~{\sc ii}] $\lambda$6548 emission lines overlap bluewards of $V_{{\rm H}\alpha}\leq-270$ km~s$^{-1}$, although [N~{\sc ii}] $\lambda$6548  is much fainter than H$\alpha$.  
Similarly the H$\alpha$ and [N~{\sc ii}] $\lambda$6584 emission lines overlap redwards of $V_{{\rm H}\alpha}\geq+535$ km~s$^{-1}$. 
Considering the contamination of the [N~{\sc ii}] emission lines to H$\alpha$, the latter is shown to extend approximately from channels centered at $V_{{\rm H}\alpha}=-632$ km~s$^{-1}$ up to $V_{{\rm H}\alpha}=+555$ km~s$^{-1}$.  
The average at $\approx-40$ km~s$^{-1}$ is consistent with the systemic velocity of $-45$ km~s$^{-1}$ reported by \citet{Gill_OBrian_2000}, whereas the maximum line-of-sight velocity $\approx590$ km~s$^{-1}$ is about 60 km~s$^{-1}$ larger than their 530 km~s$^{-1}$ estimate.


\section{Results}

The channel maps shown in Fig.~\ref{fig:channel} probe different ``velocity layers'' of the nova shell, thus providing a tomographic view \citep[e.g.\ QU\,Vul;][]{Santamaria+2022}. 
Those probing the H$\alpha$ emission line are indicative of an expanding shell; the angular size is maximum at the systemic velocity channel (between $-85$ and 0 km~s$^{-1}$) and then decreases at higher approaching and receding radial velocities. 
Meanwhile the channel maps probing the [N~{\sc ii}] $\lambda$6584 emission line follow a different pattern; the angular size is basically the same in the radial velocity range from $-200$ to $+160$ km~s$^{-1}$.  
The different structures traced by the H$\alpha$ and [N~{\sc ii}] $\lambda$6584 emission lines are discussed below using different analysis techniques.

\subsection{H$\alpha$ line maps of the ellipsoidal shell}

The H$\alpha$ line profile at each spaxel of the VLT VIMOS HR Orange data cube line maps has been fit with multiple Gaussian profiles. 
After a preliminary inspection of the spectra with the \texttt{Qfitsview} tool\footnote{https://www.mpe.mpg.de/$\sim$ott/QFitsView/}, two narrow components have been used for the nebular shell to account for its receding and approaching sides, whereas one additional broad component is required at the location of the central star. 
Single-peaked broad Balmer emission lines are typical of CVs in high-state \citep[e.g.,][]{HSK1986,DMJ1991} and can generally be attributed to the effects of disk winds \citep{MC1996}.
The wavelength range for the line peak of each component was limited up to 20~\AA\ from the blue and red nebular components to the rest frame, whereas a narrower range in wavelength of 10~\AA\ centered at the systemic wavelength was set for the stellar component. 
The line width of each nebular component was set to be larger than 4~\AA, but larger than 9~\AA\ for the stellar component. 
The best-fit parameters of each Gaussian were obtained applying a Levenberg–Marquardt least-squares fitting using an IDL-based routine. 
Examples of these fits are shown in Figure~\ref{fig:1Dfit}.

The Gaussian fits to the nebular H$\alpha$ emission line have allowed us to map the radial velocity, velocity dispersion $\sigma$, and flux in both the receding and approaching sides of the nebular shell of FH\,Ser (Fig.~\ref{fig:2Dfit}).  
These are clearly resolved within the shell (Fig.~\ref{fig:1Dfit}-middle), but blended at its edge (Fig.~\ref{fig:1Dfit}-right). 
H$\alpha$ emission is found up to 6$^{\prime\prime}$ from the central star of FH\,Ser.  
The [N\,{\sc ii}] $\lambda$6584 emission line is detected in some spaxels (as in the one shown at the middle panel of Fig.~\ref{fig:1Dfit}), but not all (as in the one shown at the right panel of Fig.~\ref{fig:1Dfit}).

The receding component is brighter towards PA$\sim$210$^\circ$, while the approaching component surface brightness peaks towards PA$\sim$60$^\circ$.  
The addition of the two components (see the bottom-left panel of Fig.~\ref{fig:img}) is consistent with a limb-brightened shell morphology. 
Both components show granularity in the flux maps (Fig.\,\ref{fig:2Dfit}-right), with clumps seen at different spatial locations. 
The blue nebular component (Fig.\,\ref{fig:2Dfit}, top-right) shows two  $\simeq 1^{\prime\prime}$ in size bright clumps towards the North and Northeast, and a fainter one towards the South, whereas the red nebular component (Fig.\,\ref{fig:2Dfit}, bottom-right) presents two clumpy arcs towards the Northeast and Southwest.  

The narrow nebular components reach velocities of $\pm$550 km~s$^{-1}$ and have similar velocity dispersion within a radius of $3^{\prime\prime}$ of the central star of $\sigma \simeq 90$ km~s$^{-1}$ (or FWHM $\simeq 210$ km~s$^{-1}$).  
Since the spectral resolution is $\simeq$100 km~s$^{-1}$ in FWHM, the lines have a width notably larger than the spectral resolution. 
The line width increases at the edge of FH\,Ser, which can be expected for a thin shell as the line of sight crosses a thicker fraction of the shell at this location, but also due to the problematic two-Gaussian fit of components too close in velocity.  
Indeed the red component shows a notorious $\sigma$ enhancement towards the South-West, reaching values of $\sigma$ up to 200~km~s$^{-1}$.

\subsection{3D visualisation of the nova remnant of FH\,Ser}

\begin{figure*}
\begin{center}
\includegraphics[width=1.0\linewidth]{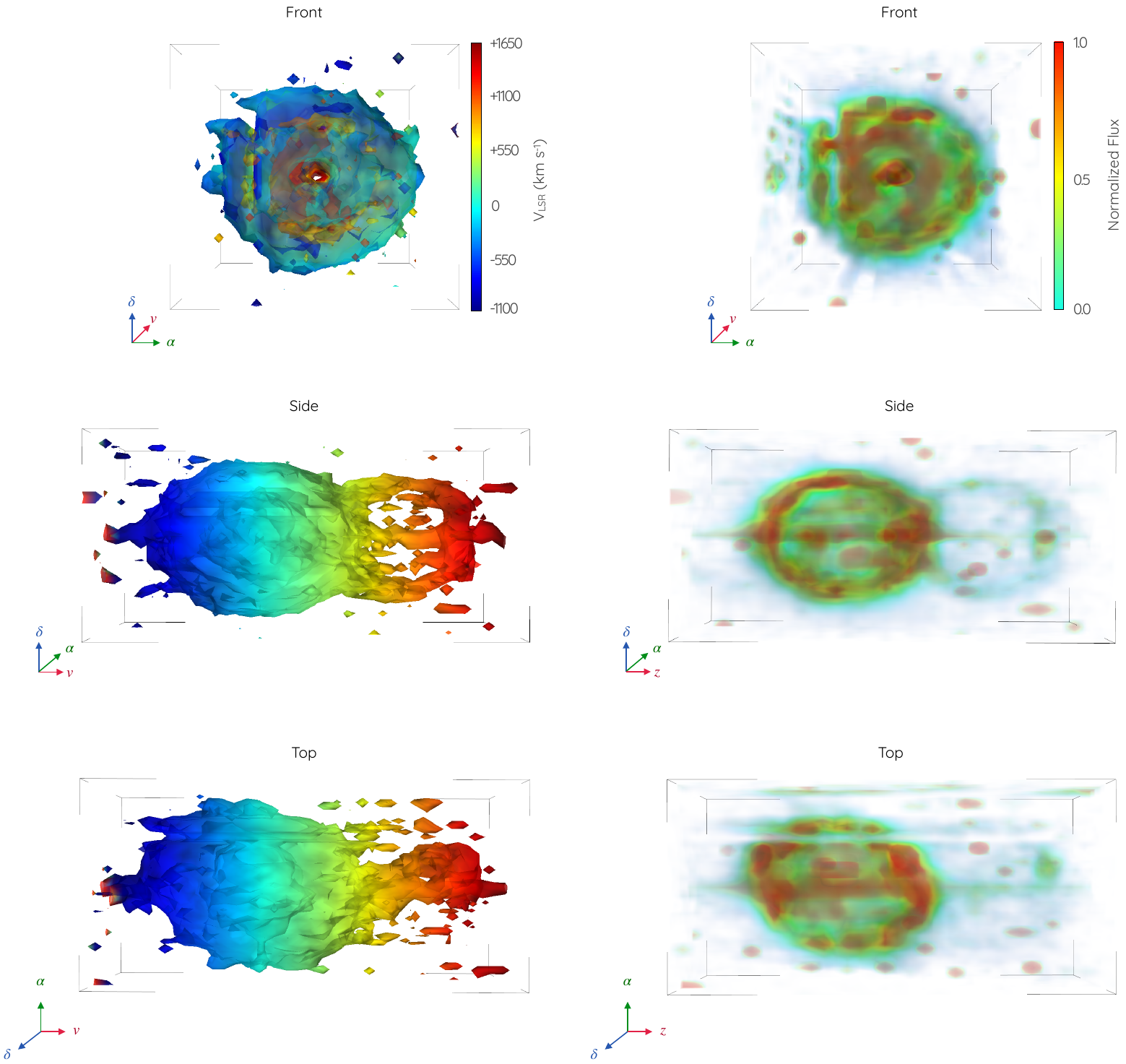}
\caption{
H$\alpha$+[N\,{\sc ii}] emission velocity-coloured (left) and intensity (right) position-position-velocity (PPV) diagrams of FH\,Ser. 
The top row shows the projection along the observer's point of view (i.e., the direct image), and the middle and bottom rows the $\delta-\lambda$ and $\alpha-\lambda$ projections from the plane of the sky along the East-West and North-South directions, respectively. 
Information on the LSR radial velocity of H$\alpha$ is provided by the rainbow colour code shown in the top-left panel, with the colour span covers the H$\alpha$ and [N~{\sc ii}] $\lambda6584$ emission lines of FH\,Ser. The colour code of the right panels corresponds to normalized flux.    
}
\label{fig:3d}
\end{center}
\end{figure*}

The 3D physical structure of FH\,Ser is clearly revealed in the H$\alpha$+[N~{\sc ii}] $\lambda\lambda$6548,6584 position-position-velocity (PPV) diagrams shown in Fig.~\ref{fig:3d} computed from the VLT VIMOS channel maps in Fig.~\ref{fig:channel} using tailor made Python routines.  
The three axes in these 3D visualisations of FH\,Ser corresponds to those of the VLT VIMOS data cube, i.e., right ascension $\alpha$, declination $\delta$, and wavelength $\lambda$ (or radial velocity).  
Three different 3D views of the H$\alpha$ and [N~{\sc ii}] $\lambda\lambda$6548,6584 emission lines are shown, the top one along the line of sight on the $\alpha-\delta$ plane of the sky (i.e., a direct image), whereas the middle and bottom correspond to the $\delta-\lambda$ and $\alpha-\lambda$ planes as observed from the plane of the sky along the East-West and North-South directions, respectively.  
The left panels display information on the radial velocity (or $\lambda$) using a rainbow colour code, whereas the right panels display information on the flux intensity. 
In these 3D visualisations, the [N~{\sc ii}] $\lambda6584$ emission line is seen as the ``red'' extension of the H$\alpha$ line, whereas the [N~{\sc ii}] $\lambda6548$ emission line is barely seen as a ``dark blue'' extension.

The 3D visualisations of the H$\alpha$ line in the middle and bottom rows of Fig.~\ref{fig:3d} confirm that its physical structure can be described by a prolate thin ellipsoid, with the expected spatio-kinematic behaviour (left panels) and limb-brightness morphology (right panels). 
This shell morphology is also seen in the [N~{\sc ii}] $\lambda6584$ emission line as the weak outermost emission in the bottom panels of Fig.~\ref{fig:3d}, particularly the right one.  
The main spatio-kinematic structure revealed by the 3D visualisations of the [N~{\sc ii}] $\lambda$6584 emission line is, however, very different compared to that of H$\alpha$. 
Rather than an ellipsoid, the [N~{\sc ii}] emission is consistent with a tilted ring seen pole on in the middle panels of Fig.~\ref{fig:3d} and sideways in the bottom panels of Fig.~\ref{fig:3d}.  
The bottom-right panel of Fig.~\ref{fig:3d} is suggestive of a similar structure in the H$\alpha$ line, but much fainter than the H$\alpha$ shell.

\subsection{3D physical structure of the nova remnant of FH\,Ser}

The H$\alpha$ channel maps in Fig.~\ref{fig:channel}, expansion velocity maps in Fig.~\ref{fig:2Dfit}, and 3D visualisations in Fig.~\ref{fig:3d} indicate that the H$\alpha$-emitting material in FH\,Ser can be mostly described as a tilted prolate ellipsoidal shell.  
If it is further assumed that its expansion is homologous, i.e., that the expansion velocity at a location of the shell is radial and proportional to the shell radius at this point, then the observed morphology and kinematics of the nova remnant depends on only a few parameters.  
Indeed a homologous expansion is supported by the constant angular expansion of the remnant in time \citep[][]{Santamaria+2020}.

Appendix~\ref{app1} describes the use of a number of spatio-kinematic measurables, including the age and H$\alpha$ radius of 4.3~arcsec along the minor axis of FH\,Ser in the VLT VIMOS images, and the maximum systemic radial expansion velocity and the radius of the location where that maximum velocity is found, to constrain the 3D physical structure of such an expanding tilted prolate ellipsoid.  
Appendix~\ref{app1} shows that FH\,Ser can be described as an expanding tilted prolate ellipsoid with ellipticity $e = 1.22\pm0.03$, equatorial and polar expansion velocity 504$\pm$27 km~s$^{-1}$ and 612$\pm$35 km~s$^{-1}$, and inclination of its major axis with the line of sight $i=52^\circ\pm7^\circ$ at a distance $950\pm50$~pc.

Appendix~\ref{app2} goes further and fits the H$\alpha$ velocity and velocity dispersion 2D maps in Fig.~\ref{fig:2Dfit} with a similar tilted prolate ellipsoid expanding homologously, but this time the shell thickness is also considered.  
This method provides rather consistent values of the  equatorial and polar expansion velocity $507\pm12$ km~s$^{-1}$ and $650\pm15$ km~s$^{-1}$, and major axis inclination $i = 52^\circ\pm2^\circ$.  
Otherwise the axial ratio $e = 1.28\pm0.02$ is a bit larger, whereas the equatorial radius $R_{\rm eq} = 3.9$ arcsec is a bit smaller.  
The distance estimate to FH\,Ser according to this method is larger, $1064^{+26}_{-23}$~pc, than in Appendix~\ref{app1}.

For comparison, \citet{Gill_OBrian_2000} concluded that FH\,Ser could be described by an ellipsoidal shell with axial ratio $1.3\pm0.1$ tilted by $62^\circ\pm4^\circ$ with the line of sight.  
Both their inclination and ellipticity or axial ratio are within the ranges of these parameters derived in Appendixes~\ref{app1} and \ref{app2}. 
Otherwise, the Gaia DR3 parallax distance \citep[1080$^{+70}_{-90}$~pc;][]{Bailer-Jones+2021} differs notably from that derived in the first method ($950\pm50$~pc), but quite similar according to the second method ($1064^{+26}_{-23}$~pc) that allows accounting for the shell thickness.

The 3D physical structure of the [N~{\sc ii}] $\lambda6584$ emission line revealed in the 3D visualisations in Fig.~\ref{fig:3d} is consistent with a ring-like or toroidal structure tilted with the line of sight.  
If this toroidal structure is assumed to be embedded within the ellipsoidal shell, as their similar spatial extent supports, its same equatorial expansion velocity of $\approx505$ km~s$^{-1}$ can be adopted.  
Then the observed maximum expansion velocity $\approx\pm435$ km~s$^{-1}$ at the tips of the toroidal structure imply that it is tilted by $\simeq$49$^\circ$ with the plane of the sky, which is consistent with that of the ellipsoidal shell.  
The [N~{\sc ii}] toroidal structure thus forms a belt at the equator of the H$\alpha$ ellipsoidal shell.

\subsection{Ionized mass and its kinetic energy}

The ionized mass $M_\mathrm{ion}$ of FH\,Ser can be derived from its intrinsic H$\alpha$ flux as \citep[see][]{1968IAUS...34..162B} 
\begin{equation}
    M_\mathrm{ion} = \mu m_{\rm p}N_{\rm p}V\varepsilon
\end{equation}
where $\mu$ is the mean molecular weight, which can be assumed to be 1.44 for a He/H solar ratio, m$_{\mathrm p}$ is the proton mass, $N_{\mathrm p}$ is the proton density, $V$ is the emitting volume, and $\varepsilon$ is the filling factor, which can be described as $\varepsilon = a \times b$, where $a$ and $b$ are the macroscopic and microscopic filling factor components, respectively \citep[see details in][]{Santamaria+2022}.  
This can be expressed as 
\begin{equation}
    M_\mathrm{ion} = \mu m_{\rm p} \sqrt{\frac{4 \pi d^2 \varepsilon V F_\mathrm{H\alpha}}{j_\mathrm{H\alpha}}}
\end{equation}
where $d$ is the distance to FH\,Ser, $F_{{\mathrm H}\alpha}$ is the unabsorbed (intrinsic) H$\alpha$ flux, and $j_\mathrm{H\alpha}$ is the emission coefficient of the H$\alpha$ line with a value of $4\times10^{-25}$ erg~cm$^{-3}$~s$^{-1}$ \citep{OF2006}. 
Similarly the root mean square (rms) density $n_\mathrm{e}$ can be expressed as 
\begin{equation}
    n_\mathrm{e} = \sqrt{\frac{4 \pi d^2 \mu F_\mathrm{H\alpha}}{\varepsilon V j_\mathrm{H\alpha}}}. 
\end{equation}

The total observed H$\alpha$ flux derived from the VLT VIMOS IFS observations is $2.7\times10^{-14}$ erg cm$^{-2}$ s$^{-1}$.  
The extinction towards FH\,Ser can be estimated using the reddening versus distance curve along its direction provided by Bayestar19\footnote{http://argonaut.skymaps.info} \citep{Green+2018}.  
At the Gaia distance of 1080 pc for FH Ser, the extinction is quite flat, with $E(g-r) = 0.53\pm0.01$ mag in the distance range from 730 to 1440 pc.  
The conversion from $E(g-r)$ to $E(B-V)$ provided by \citet{SF2011} implies $c$(H$\beta$)=0.77 or $E(B-V)=0.54$ mag. 
After applying the extinction correction, the intrinsic H$\alpha$ flux is estimated to be $F_\mathrm{H\alpha}$=8.8$\times10^{-14}$ erg cm$^{-2}$ s$^{-1}$.
This results in a total ionised mass $M_\mathrm{ion}=4.6\times10^{-4} \times \varepsilon^{1/2}$ M$_\odot$, and an rms electron density $n_\mathrm{e}$ of $\simeq240\times \varepsilon^{-1/2}$ cm$^{-3}$.

Actually, IFS observations allow us the chance to evaluate the ionized mass on each volume element of the 3D datacube once the velocity (or wavelength) along the $z$ axis can be converted into depth in physical scale (cm, pc, or arcsec). 
This way information on the true 3D structure and the inhomogeneous clumpy distribution of the ejected material in the nebular shell can be taken into account, reducing the uncertainty on the filling factor $\varepsilon$ by removing its macroscopic term $a$. 
The equatorial expansion velocity of FH\,Ser is $\simeq$505 km~s$^{-1}$, which corresponds to an equatorial radius of 4.3 arcsec, or $\simeq7\times10^{16}$ cm at the Gaia DR3 distance of 1080 pc. 
Thus the scaling factor between velocity and linear radius  
\begin{equation}
    r\,(\mathrm{cm}) = 1.38\times10^{14} \; V\,(\mathrm{km~s}^{-1}), 
\end{equation}
implies that the spectral dispersion of 0.6 \AA~pix$^{-1}$ (or 27.4 km~s$^{-1}$~pix$^{-1}$ at H$\alpha$) and plate scale of $0.66\times0.66$ arcsec$^2$~pix$^{-1}$ corresponds to a volume element of the 3D datacube of $\simeq4.3\times10^{47}$~cm$^3$. 

The total mass of the nebula, as estimated for each volume element of the datacube and them added together, results to be $2.6\times10^{-4} \times b^{1/2}$ M$_\odot$. 
A comparison of this mass and the mass derived from the total H$\alpha$ flux computed above implies a value of 0.7 for $a$, the macroscopic component of the filling factor.

The total volume occupied by the nova shell implies a swept-up mass of $2.4\times10^{-6}$ M$_\odot$ for an assumed value of 1 cm$^{-1}$ for the density of the circumstellar medium. 
The mass of the nova ejecta is thus much greater than the mass of the swept ISM, by a factor $188\times b^{1/2}$.  
This is fully consistent with a free expansion phase, as confirmed by the constant angular expansion of the shell.  
This is supported as well by the kinetic energy of FH\,Ser, $E_{\rm kin} = 1.6\times10^{45}$ erg, which has been derived by adopting an expansion velocity weighted between the radial velocity and the velocity on the plane of the sky.

\section{Discussion}

The IFS observations of the nova remnant FH\,Ser presented here disclose two distinct structural components:
a prolate ellipsoidal shell most prominent in H$\alpha$ (and very weak in [N~{\sc ii}]) and an equatorial belt best seen in [N~{\sc ii}] (and probably also present in H$\alpha$).  
The earliest evidence of these structures was actually presented by \citet{Hutchings1972}, who modeled the H$\delta$ line profile using a spherically symmetric shell and a polar-cap shell.  
Similar investigation on the early development of asymmetries in its ejecta was carried out by \citet{SP1977} based on radio observations.

There is a growing number of nova remnants that display different large-scale structural components distinctly revealed in H~{\sc i} Balmer emission lines and forbidden [N~{\sc ii}] and [O~{\sc iii}] emission lines, e.g., HR\,Del \citep{HO2003,MD2009}, Sab\,142 \citep[a.k.a.\ IPHASX\,J210204.7$+$471015,][]{Guerrero+2018}, T\,Pyx \citep{Izzo+2024,Santamaria+2024}, and RR\,Pic \citep{Celedon+2024}. 
These different structural components most likely inform us about chemical or density inhomogeneities or varying excitation conditions in the nova ejecta formed either during the TNR, or resulting from the interaction of the nova ejecta with the accretion disk and donor companion \citep{DO2010}.  
Indeed these are quite notorious in early near-IR IFS observations of nova remnants \citep[e.g., V723\,Cas, a.ka.a.\ Nova Cas 1995, and V5668\,Sgr, a.k.a.\ Nova Sgr 2015b;][]{LC2009,Takeda+2022}.
The rather similar inclination of the symmetry axes of both structures with respect to the line of sight in FH\,Ser suggests that they both formed as the result of the same shaping process.  
The most likely explanation is the interaction of the nova ejecta with the accretion disk around the WD\footnote{Note that the accretion disk is not necessarily coplanar with the orbital plane, but a tilted precessing disk is very frequent in CVs as revealed by superhumps in their light curves \citep[e.g., BK\,Lyn;][]{Patterson+2013}. 
}
that strips material from the disk, which is not completely disrupted, and mass-load the ejecta preferentially along the plane of the accretion disk \citep{Booth+2016,Figueira+2018}, slowing it down. 
The interaction of these ejecta with the subsequent fast stellar wind from the WD will result in the ellisoidal shell morphology \citep[e.g.,][]{Orlando2017}, as the latter encounters higher density and slower expanding material along the accretion disk plane. 
The mass-loading of the ejecta is also supported by the ionized mass of FH\,Ser, much larger than the theoretical predictions for nova outburst of similar expansion velocity and decline time \citep{Yaron+2005}.

\citet{Celedon+2024} argued that equatorial structures most prominent in forbidden lines have to be associated with the lower density material of evolved, old nova remnants, but the young age of FH\,Ser (39.37 yrs at the time of the VLT VIMOS observations) contradicts this.  
It is more likely that the interaction of the ejecta with the accretion disk enhances shocks that amplify the emission in shock-sensitive emission lines such those of [N~{\sc ii}].  
This is also the case of Sab\,142 \citep{Guerrero+2018}, where the H~{\sc i} Balmer lines are not even detected in the equatorial ejecta, but contrary to T\,Pyx \citep{Izzo+2024,Santamaria+2024}, where the equatorial ring-like structure is most prominent in the H~{\sc i} Balmer lines. 
These differences highlight the diversity of nova remnants and complexity of their shaping.

It is worth noting that narrow-band H$\alpha$ or [N~{\sc ii}] images of nova remnants can suffer the contamination of one emission line to the image of the other given their high expansion velocities. 
This is an important issue particularly for nova remnants with different structural components traced by the H$\alpha$ and [N~{\sc ii}] emissions. 
Thus structural features associated with one emission line may show in the narrow-band image of the other emission line. 
Indeed the HST WFPC2 F656N filter used to obtain the image shown in the top-right panel of Fig.~\ref{fig:img} has a pivotal wavelength of 6563.81 \AA\ and a FWHM of 28.55 \AA, resulting in a non-negligible sensitivity to the total emission of the [N~{\sc ii}] $\lambda\lambda$6548,6584 emission lines.  
It is thus very likely that the equatorial ring in this image includes a significant contribution of [N~{\sc ii}] emission. 
The images of HR\,Del \citep{HO2003} and T\,Pyx \citep{Izzo+2024,Santamaria+2024} are also most likely affected by this issue.  
This emphasizes the advantages (even the need) of using IFS observations rather than direct imaging in order to reveal the true morphology and physical structure of nova remnants.

The use of IFS observations provides a critical advantage in fitting the geometry and inclination of an ellipsoidal shell with the homologous expansion describing the nova remnant.  
Two different methods have been developed here to obtain best-fit parameters of this ellipsoidal shell, one in Appendix~\ref{app1} using spatio-kinematic information along the major axis of FH\,Ser, and a second one in Appendix~\ref{app2} using 2D velocity maps that can be uniquely derived from IFS observations.  
Both methods provide similar (although not exactly the same) best-fit parameters.  
One major difference results for the expansion distance to FH\,Ser, which is discrepant with respect to the Gaia DR3 parallax distance \citep[1080$^{+70}_{-90}$ pc;][]{Bailer-Jones+2021} in the first method, but quite similar according to the second method that allows accounting for the shell thickness.  
It is well-known that the Gaia determination of parallax distances to nova remnants results in notably different distances than those derived from their nebular expansion, very particularly for sources farther than 1.0 kpc \citep{Schaefer2018,dVI2020}.  
This discrepancy can be attributed to the methods used to derive distances from the Gaia's parallax, as discussed in length by \citep{SG2019} and \citet{Tappert+2020}, but also to difficulties in the expansion distance method for barely resolved nova remnants \citep{DD2000} or by its application to expanding prolate ellisoidal shells \citep{WHC2000}.  
The method described in Appendix~\ref{app2} shows that the consideration of the shell thickness in the determination of the expansion distance fixes the problem of a discrepant Gaia's distance for FH\,Ser.  
We have started a large program to fit available IFS observations of nova remnants using the full potential of these spatio-kinematic data, in conjunction with available narrow-band images to further use their spatial information to better constrain the models.


\section{Conclusions}

FH\,Ser is a slow classical nova that experienced a nova outburst in February 1970. 
Its nova remnant is elliptical in shape, with multiple knots, and a peculiar ring-like filament along its minor axis.
Integral field spectroscopic VLT VIMOS observations with a spectral resolution $R\sim3000$ of the nebular remnant of FH\,Ser have been used to investigate its spatio-kinematic structure in the H$\alpha$ and [N~{\sc ii}] 6548,6584 emission lines.

The data cube has been analyzed using position-position-velocity (PPV) diagrams of its velocity field and emission intensity.  
The 3D visualisations of FH\,Ser uniquely reveal the presence of two different structural components, a tilted prolate ellisoidal shell, most prominent in H$\alpha$, and a ring-like structure, most prominent in the forbidden emission line of [N~{\sc ii}]. 
The best H$\alpha$ narrow-band images of FH\,Ser do actually mix the emission from these two components, with the outer elliptical envelope dominated by the H$\alpha$ emission and the knotty and filamentary inner ring-like structure dominated by [N~{\sc ii}] emission.

The spatio-kinematics of the H$\alpha$ emission has been modeled adopting a prolate ellipsoidal shell model with a homologous expansion.  
The data have been fit using two different techniques to determine the inclination, axial ratio, and polar and equatorial velocity of such prolate ellipsoid.  
The first one uses a number of particular measurables of a tilted expanding ellipsoid to derive these parameters, whereas the second one takes full advantage of the integral field spectroscopic observations to model the 2D velocity maps using a Markov Chain Montecarlo fit.  
Both fits imply an inclination of the major axis of the ellipsoid of $\approx$52$^\circ$ with the line of sight, a major-to-minor axial ratio of $\approx$1.25, and an equatorial velocity of $\approx$505 km~s$^{-1}$. 
The relatively slow expansion velocity and large axial ratio agrees with the classification of FH\,Ser as a slow nova.

The [N~{\sc ii}]-dominated ring-like structure shares the same inclination angle than the H$\alpha$-dominated ellipsoidal shell. 
It can thus be described as an equatorial belt-like structure of the ellipsoidal shell that can be linked to the process that shaped the latter.  
The most likely connection is the interaction of the nova ejecta with the accretion disk around the WD at the time of the nova outburst.  
Stripped material from the accretion disk is proposed to mass-load the nova ejecta, slowing it down on the equatorial plane.

The 3D structure of FH\,Ser has been used to refine the estimate of its ionized mass, found to be $M_\mathrm{ion}=2.6\times10^{-4}$ M$_\odot$.  
The ionized mass is much larger than the interstellar medium swept up by the nova shell, which conforms with its free expansion and high kinetic energy of $1.6\times10^{45}$ erg.  
It is also much larger than the theoretical expectations for the ejecta of a slow nova, confirming mass-loading of the nova ejecta.

This work demonstrates that integral field spectroscopy at a dispersion $R \gtrsim 3000$ provides suitable means to separate different structural components of nova remnants, to derive reliable estimates of key parameters, such as the shell aspect ratio, its ionized mass, and expansion distance, and to spatio-kinematically resolve small-scale structures (clumping).  
New methods to analyze the rich data cube provided by integral field spectroscopic observations of the complex nova remnants are being developed for these purposes.


\begin{acknowledgements}

M.A.G.\ and S.C.\ acknowledge financial support from grant CEX2021-001131-S funded by MCIN/AEI/10.13039/501100011033. 
M.A.G.\ also acknowledges financial support from grant PID2022-142925NB-I00 from the Spanish Ministerio de Ciencia, Innovaci\'on y Universidades (MCIU) cofunded with FEDER funds. 
E.S.\ thanks UNAM DGAPA for a postdoctoral fellowship.
J.A.T. and E.S.\ acknowledges support from the UNAM PAPIIT project IN~102324. 
This work has made extensive use of NASA's Astrophysics Data System (ADS).
A.E.\ acknowledges the financial support from the Spanish Ministry of Science and Innovation and the European Union - NextGenerationEU through the Recovery and Resilience Facility project ICTS-MRR-2021-03-CEFCA .

\end{acknowledgements}


%
%




\appendix

\section{One-dimensional analytical fit of a homologous expanding prolate ellipsoidal shell}
\label{app1}

The observed morphology and kinematics of the nova remnant of FH\,Ser, on the assumption that it can be described by a tilted prolate ellipsoidal shell with homologous expansion, depend only on a few parameters, namely (1) the shell aspect ratio $e$ between the semi-major axis $a$ and the semi-minor axis $b$, 
\begin{equation}
    e = a/b, 
\end{equation}
(2) the inclination of the major-axis with the line of sight $i$, 
(3) the distance $d$, 
(4) the semi-minor axis $b$, and 
(5) the remnant age $\Delta t$. 
The latter two parameters can be fixed to the time of the VLT VIMOS observations (i.e., $b = 4.3$ arcsec and $\Delta t = 39.37$ yr).

Other key parameters, such as the equatorial and polar expansion velocities, $V_{\rm eq}$ and $V_{\rm p}$, respectively, can be simply expressed in terms of the distance and aspect ratio, given the known remnant age and semi-minor axis, as 
\begin{equation}
    V_{\rm eq} \; {\rm (km~s}^{-1}{\rm )} = 0.53 \times 
    d \; {\rm (pc)},
\label{eq.app.veq}
\end{equation}
which is just another way of writing the angular expansion along the minor axis derived by \citet{Santamaria+2020} of $\dot\theta = 0\farcs109$~yr$^{-1}$, and 
\begin{equation}
    V_{\rm p} \; {\rm (km~s}^{-1}{\rm )} = 0.53 \times 
    e \times d \; {\rm (pc)}. 
\label{eq.app.vp}
\end{equation}

\subsection{Special measurables of a tilted expanding ellipsoid}

Let's consider Figure~\ref{fig:app.sketch} as a description of the slice of an ellipsoid along its projected major axis.  
There are a number of spatio-kinematic measurables in a tilted expanding ellipsoid that can be used to determine the unknown parameters $e$, $i$, and $d$.  
The relationships between these spatio-kinematic measurables and the unknown parameters are described next.

\subsubsection{Projected semi-major axis}

\begin{figure}
    \begin{center}
\includegraphics[bb=60 40 553 600,width=0.9\linewidth]{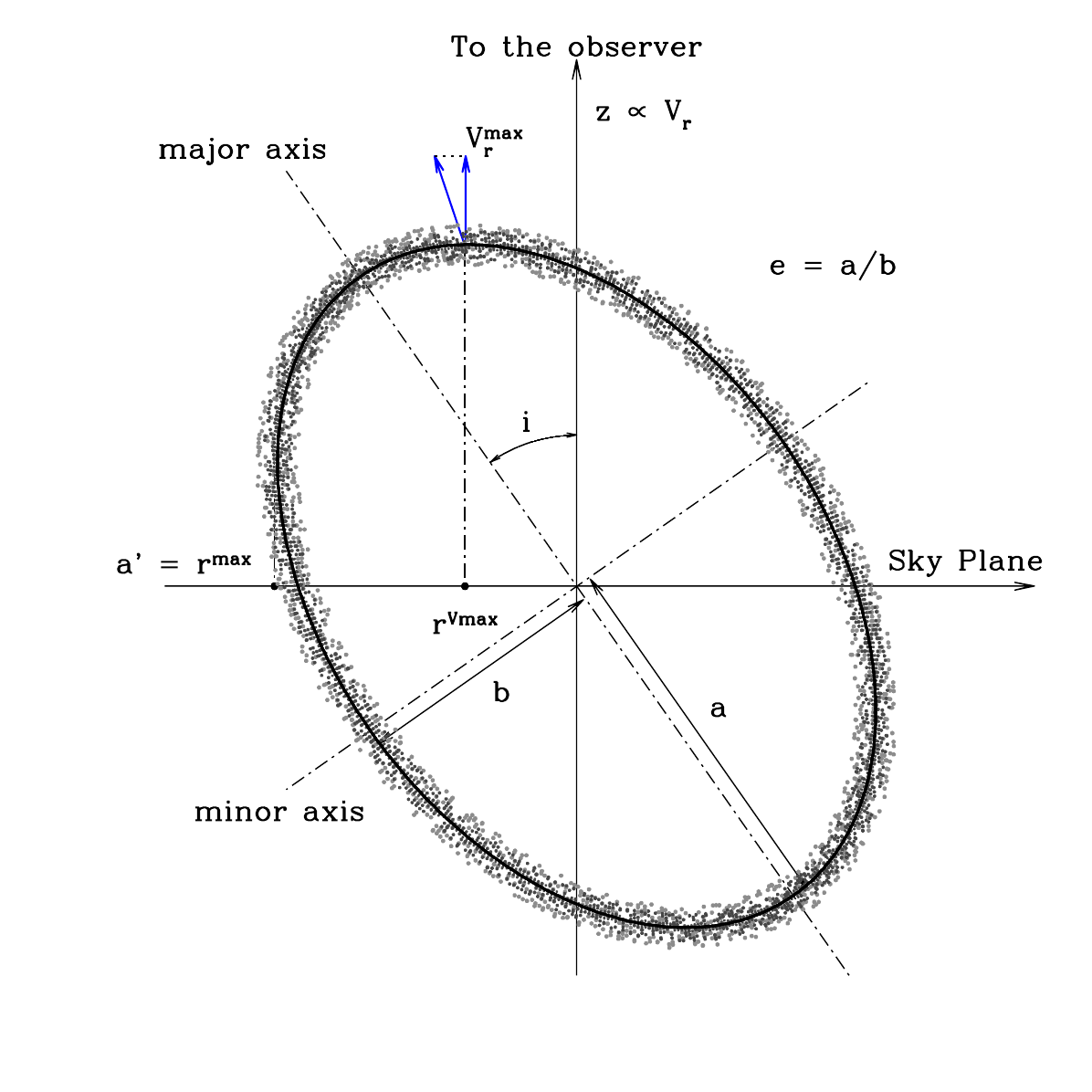}
\caption{
Sketch of a slice of the ellipsoidal shell model across the plane including its major axis and tilt direction. 
The geometry of the ellipsoid is defined by the semi-major axis $a$, semi-minor axis $b$, their ratio $e$, and the inclination angle $i$ with the line of sight, illustrated on the sketch. 
The singular measurables projected semi-major axis $r^\mathrm{max}$, maximum systemic radial velocity $V^\mathrm{max}_\mathrm{r}$, and radial distance at maximum systemic radial velocity $r^{V^\mathrm{max}_\mathrm{r}}$ are also labeled. 
}
\label{fig:app.sketch}
\end{center}
\end{figure}

The Cartesian coordinates of the ellipse representative of the slice along the major axis of an ellipsoid with semi-minor axis $b$ and aspect ratio $e$ can be written as 
\begin{equation}
    r = b \; \cos \phi; \;\;\; z = a \; \sin \phi = e \; b \; sin \phi, 
\end{equation}
which is the thin ellipse shown in Figure~\ref{fig:app.sketch}, where $r$ is the distance on the plane of the sky to the center, $z$ is the distance to the plane of the sky, and $\phi$ is the angle with the semi-major axis, which varies from $0^\circ$ to $360^\circ$.  
Considering an inclination angle of the major-axis with respect to the line of sight $i$, then the value $b=4.3$ arcsec and the relationship $e=a/b$ can be used to derive the projected radius of a point on this tilted ellipse  
\begin{equation}
    r' = 4.3 \times ( \cos \phi \cos i \; + \; e \;  \sin \phi \sin i) \;\; {\rm {(arcsec)}}
\end{equation}
and its distance to the plane of the sky
\begin{equation}
    z' = 4.3 \times (-\cos \phi \sin i \; + \; e \;  \sin \phi \cos i) \;\; {\rm {(arcsec)}}. 
\end{equation}
This tilted ellipse is shown in Figure~\ref{fig:app.sketch} by a thick ellipse, where the grey dots around it remind us that the observed remnant shell is not infinitely thin, but a distribution of emitting volumes distributed within a relatively thin shell.

At the maximum projected distance 
\begin{equation}
    \frac{dr'}{d\phi} = 0 \; \rightarrow  \; \sin \phi \cos i \; = e \; \cos \phi \sin i \; \leftrightarrow \; \tan \phi = e \; \tan i, 
\end{equation}
Expressing $\cos \phi$ and $\sin \phi$ as a function of $\tan i$, it can be shown that
\begin{equation}
    r_{\rm max} = b \; (A^{1/2} \cos i + e (1-A)^{1/2} \sin i), 
    \label{eq.app1.rmax1}
\end{equation}
where
\begin{equation}
    A = \frac{1}{1+e^2\tan^2 i} = \cos^2 \phi
    \label{eq.app1.rmax2}
\end{equation}
for $\phi$ at the maximum radial projection on the sky. 
The value of $r_{\rm max}$ is estimated to be 4.9 arcsec, thus the projected (observed) axial ratio is $\simeq 1.14\pm0.03$, adopting an uncertainty of 0.1 arcsec for $r_{\rm max}$ and $b$.

\begin{figure}
    \begin{center}
\includegraphics[bb=20 30 503 500,width=0.90\linewidth]{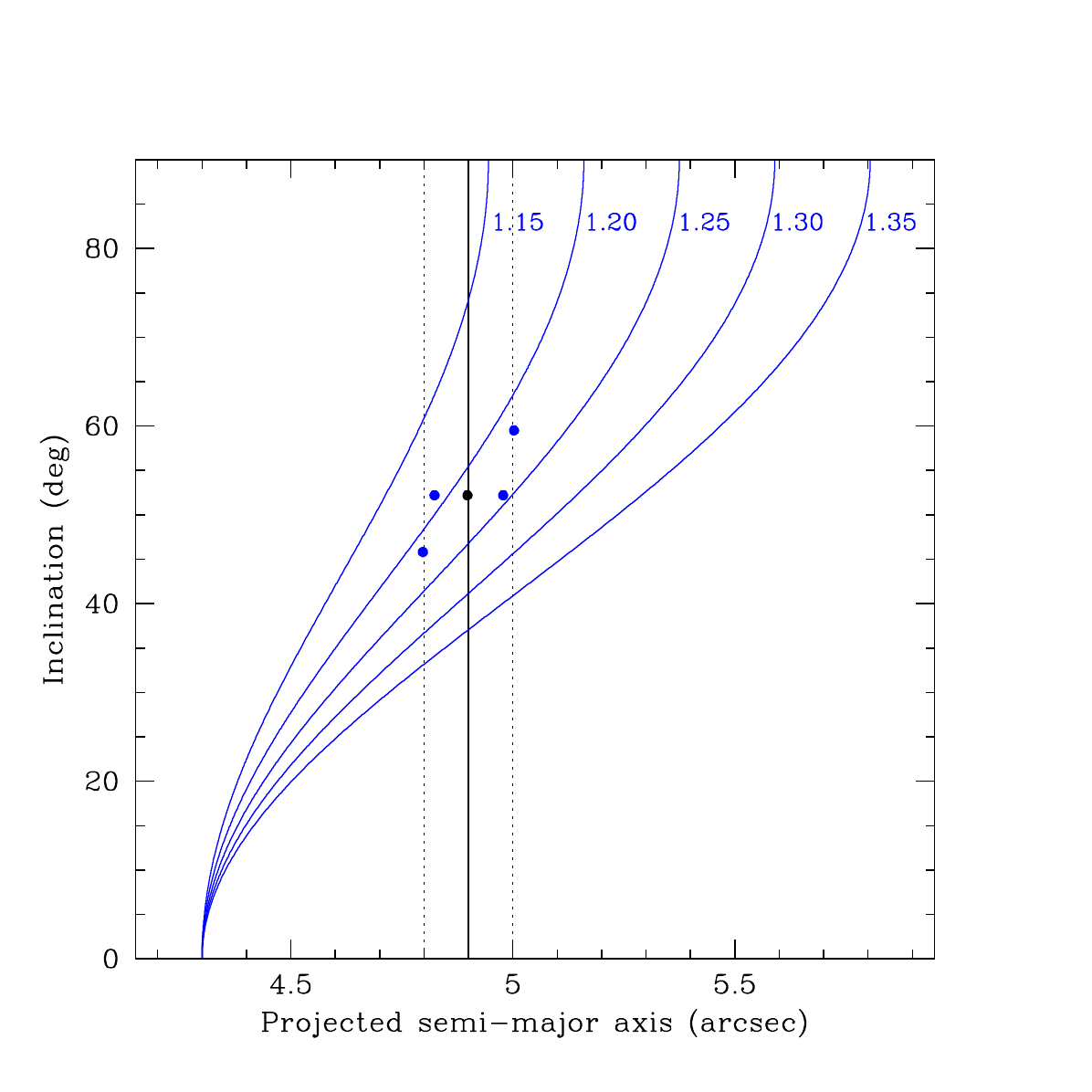}
\caption{
Variation of the projected semi-major axis of an ellipsoid along its symmetry axis for 5 different values of its ellipticity (labeled in blue) with its major axis inclination. 
The observed semi-major axis of 4\farcs9 of FH\,Ser is marked by a solid vertical line, with 1-$\sigma$ uncertainty marked by vertical dotted lines. 
The black dot corresponds to the best fit value, with blue dots in the left and right panels representing acceptable values within the uncertainty of the observed parameters.
}
\label{fig:app.xmax}
\end{center}
\end{figure}

Figure~\ref{fig:app.xmax} shows the variation of the observed axial ratio with the inclination angle for 5 different values of the true ellipticity.  
As expected, the observed axial ratio increases with the true axial ratio and the tilt of the major axis with the line of sight.

\subsubsection{Maximum systemic radial velocity}

The systemic radial velocity at any point of an ellipsoid with homologous expansion is proportional to the distance to the plane of the sky $z$.  
Thus it can be expressed as 
\begin{equation}
    V_{\rm r} = 0.53 \times d {\rm {(pc)}} \; (-\cos \phi \sin i \; + \; e \;  \sin \phi \cos i) \;\; {\rm (km~s}^{-1}{\rm )}. 
\end{equation}
As in the previous section, the maximum systemic radial velocity can be computed as 
\begin{equation}
    \frac{dV_{\rm r}}{d\phi} = 0 \; \rightarrow  \; \sin \phi \sin i \; = - e \; \cos \phi \cos i \; \leftrightarrow \; \tan \phi = \frac{-e}{\tan i}, 
\end{equation}
that after expressing $\cos \phi$ and $\sin \phi$ as a function of $\tan i$, results in the following expression
\begin{equation}
    \frac{V_{\rm r}^{\rm max}}{d {\rm {(pc)}}} = 0.53 \times (-B^{1/2} \sin i + e (1-B)^{1/2} \cos i), 
\end{equation}
where
\begin{equation}
    B = \frac{1}{1+\frac{e^2}{\tan^2 i}} = \cos^2 \phi
\end{equation}
for $\phi$ at the maximum systemic radial velocity.

\begin{figure}
    \begin{center}
\includegraphics[bb=20 30 503 500,width=0.90\linewidth]{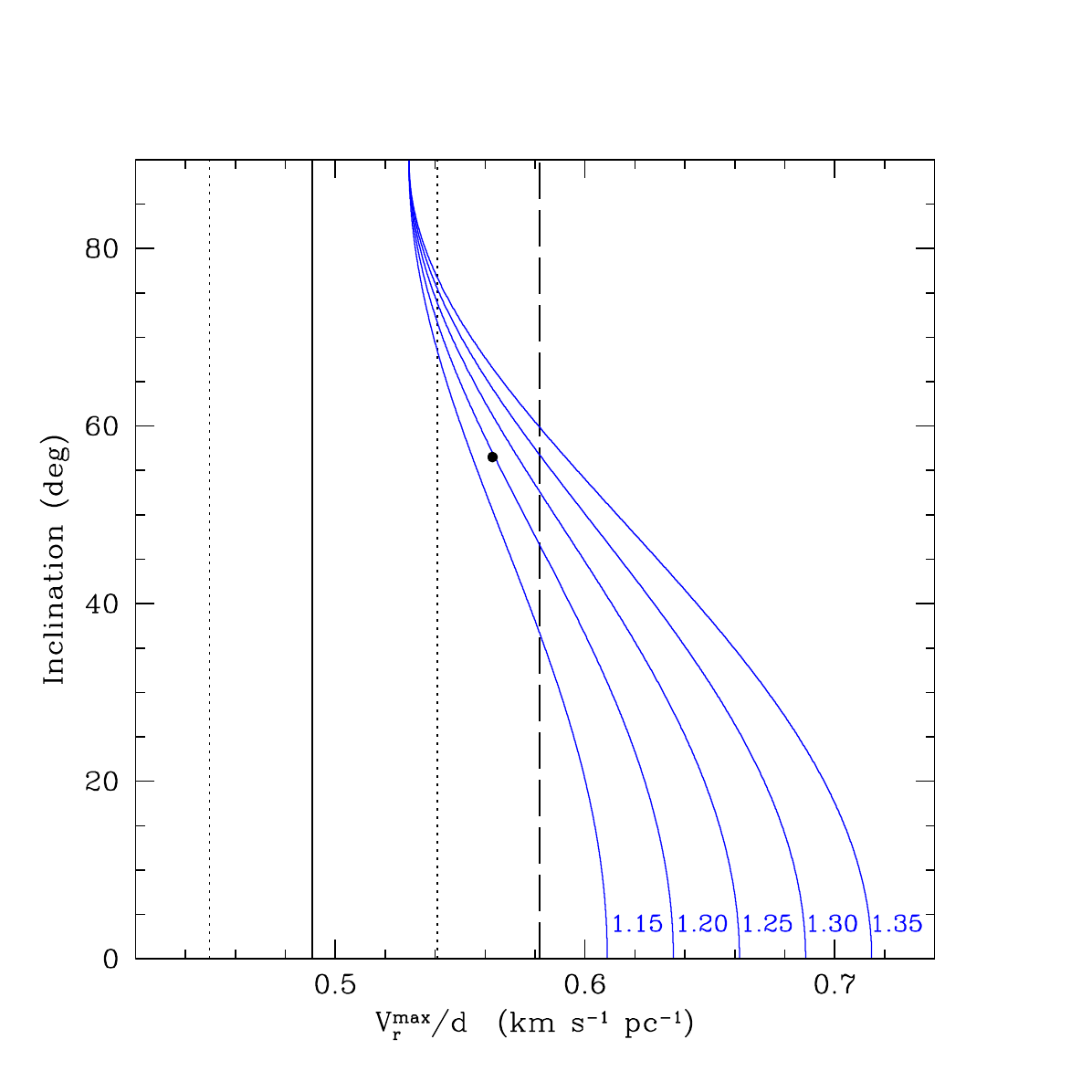}
\caption{
Variation of the ratio of the maximum systemic radial velocity to the distance of an ellipsoid for 5 different values of its ellipticity (labeled in blue) with its major axis inclination. 
The ratio between the observed maximum systemic radial velocity of FH\,Ser and its \emph{Gaia} distance is marked by a solid vertical line, with 1-$\sigma$ uncertainty marked by vertical dotted lines and the 2-$\sigma$ lower uncertainty by a vertical dashed line. 
The black dot corresponds to the best fit value. 
}
\label{fig:app.vmax}
\end{center}
\end{figure}

Figure~\ref{fig:app.vmax} shows the variation of the ratio of the observed maximum systemic radial velocity and distance with the inclination angle for 5 different values of the true ellipticity.  
The observed maximum systemic radial velocity is estimated to be $530\pm10$ km~s$^{-1}$, whereas the \emph{Gaia} distance to FH\,Ser is $1080^{+70}_{-90}$ pc.  
The maximum systemic radial velocity is shown to increase with axial ratio and decrease with increasing tilt of the ellipsoid with the line of sight.  
We notice that there is marginal agreement between the predicted and expected values of this ratio, suggesting that FH\,Ser could be located closer than indicated by its \emph{Gaia} distance.  
The 2-$\sigma$ lower uncertainty in the \emph{Gaia} distance is shown in this plot as an additional vertical dashed line.

\subsubsection{Radial distance at maximum systemic radial velocity}

\begin{figure}
    \begin{center}
\includegraphics[bb=20 30 503 500,width=0.90\linewidth]{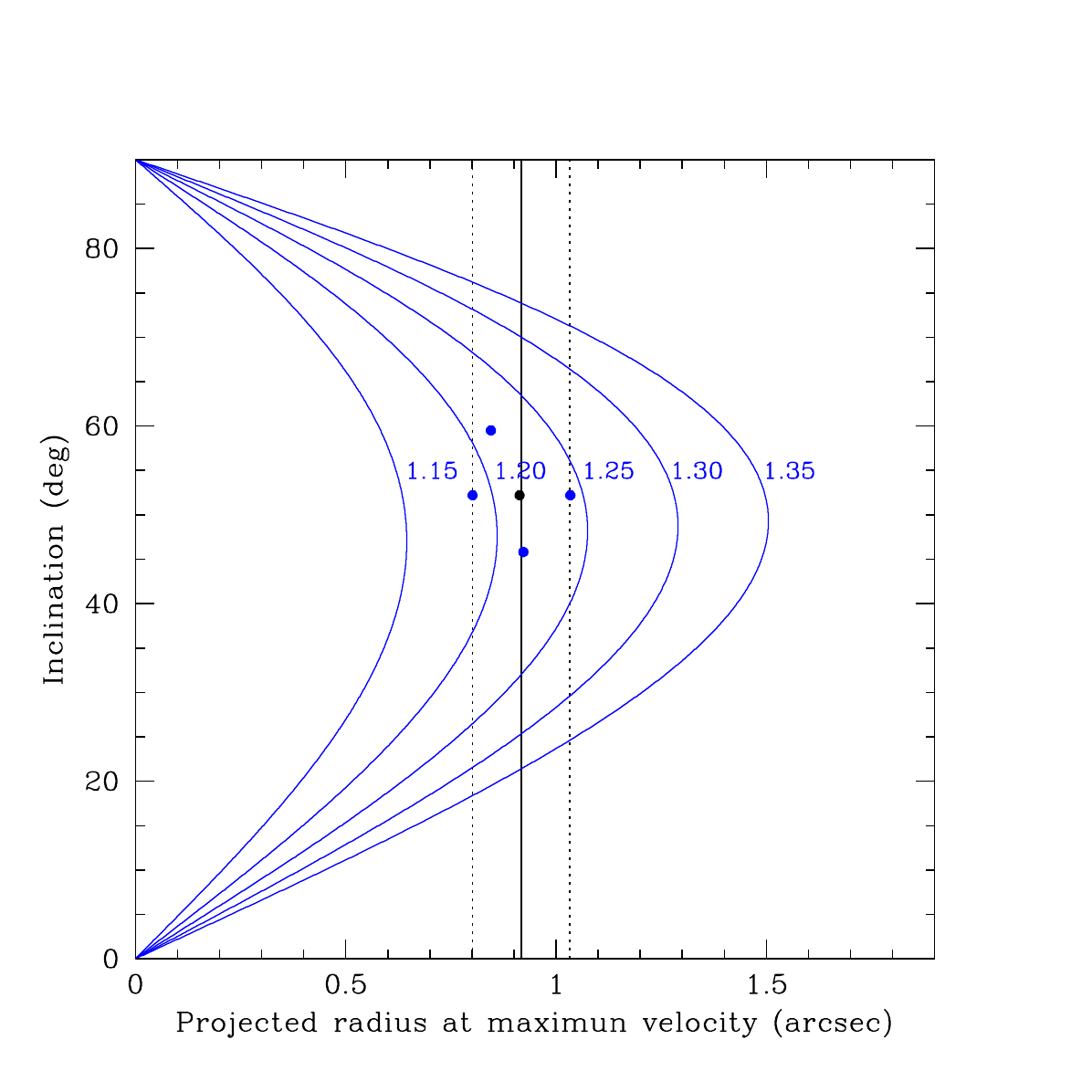}
\caption{
Variation of the projected radius at the maximum systemic radial velocity for 5 different values of its ellipticity (labeled in blue) with its major axis inclination. 
The observed projected radius at the maximum systemic radial velocity of FH\,Ser is marked by a solid vertical line, with 1-$\sigma$ uncertainty marked by vertical dotted lines. 
In all panels, the black dot corresponds to the best fit value, with blue dots in the left and right panels representing acceptable values within the uncertainty of the observed parameters. 
}
\label{fig:app.xvmax}
\end{center}
\end{figure}

The above maximum systemic radial velocity is achieved at a projected radius
\begin{equation}
    r^{\rm Vmax} = b \; (-B^{1/2} \cos i + e (1-B)^{1/2} \sin i), 
\end{equation}
which is plotted in the right panel of Figure~\ref{fig:app.xvmax} for 5 different values of the true ellipticity. 
The projected radius of the point with the maximum systemic radial velocity is measured in the velocity maps presented in Figure~\ref{fig:2Dfit}.  
This results to be $0.92\pm0.12$ arcsec, which is plotted in Figure~\ref{fig:app.xvmax} as a vertical solid line with 1-$\sigma$ uncertainty as dotted lines. 

%
%

\subsection{Geometry of the nova remnant of FH\,Ser}

Figures~\ref{fig:app.xmax}, \ref{fig:app.vmax}, and \ref{fig:app.xvmax} place some constraints on the different parameters $e$, $i$, and $d$. 
The plots of projected axial ratio (Fig.~\ref{fig:app.xmax}) and projected point at maximum systemic radial velocity (Fig.~\ref{fig:app.xvmax}) are clearly contrasting.  
The best joint-fit to both variations is achieved only for a relatively low ellipticity, $\simeq1.215$, and an intermediate inclination angle, $i\simeq52.2^\circ$ (black dot in these figures).  
Accounting for the 1-$\sigma$ uncertainty, fits can be achieved within $e$ and $i$ ranges $1.189 \leq e \leq 1.243$ and $45.8^\circ \leq i \leq 59.5^\circ$ (blue dots in these figures).  
Therefore the values of $e$ and $i$ are estimated to be 1.215$\pm$0.027 and $52^\circ\pm7^\circ$, respectively.

On the other hand, the ratio of the maximum systemic radial velocity to distance (Fig.~\ref{fig:app.vmax}) is inconsistent with the geometrical models, unless the distance is notably lower, 950$\pm$50 pc, than the \emph{Gaia} distance of $1080^{+70}_{-90}$ pc.  
The equatorial and polar velocity would be 504$\pm$27 and 612$\pm$35 km~s$^{-1}$, respectively.

\section{Two-dimensional Montecarlo model fitting of a homologous expanding prolate ellipsoidal shell}
\label{app2}

The velocity and velocity dispersion maps (left and middle panels of Fig.~\ref{fig:2Dfit}, respectively) provide means for a robust, 2D fit of the expanding shell of FH\,Ser.  
The fits is described below together with the results of these fits.

\subsection{Geometrical model}

The same tilted prolate ellipsoidal shell model with homologous expansion (see Fig~\ref{fig:app.sketch}) is adopted, thus its morphology depends only on two free parameters,  
(1) the mean equatorial radius $R_{\rm{eq}}\in\mathbb{R}^{+}$, which corresponds to the semi-minor axis, $b$, and 
(2) the shell aspect ratio $e$ between the semi-major axis, $a$, and the semi-minor axis, $b$, i.e, $e=a/b\in[1,\infty)$.  
Unlike in Appendix~\ref{app1}, this ellipsoid will be assumed here to have a finite thickness following a normal distribution with standard deviation $w$ from its central value.  
The value of $w$ has been fixed to 0.13 arcsec, according to the FWHM of 0.31 arcsec of the shell measured in HST images.

The model will additionally have three internal model parameters; 
the azimuthal angle $\theta\in(-\pi,\pi]$ measured from the $x$ axis (the North direction on the plane of the sky), 
the polar angle $\phi\in[0,\pi]$ measured from the $z$ axis (the line of sight), and 
the depth parameter $\xi\in [-3,+3]$ that, for given values of the polar and azimuthal angles, sets the ``distance'' of a point within the shell from the mean radius along this direction.

The model of a prolate spherical shell is thus described by the following equations
\begin{equation}\label{eq:app2.basic_model}
\left\{
\begin{array}{l}
x = (R_{\rm{eq}} + w\xi)\cos\theta\sin\phi \\
y = (R_{\rm{eq}} + w\xi)\sin\theta\sin\phi \\
z = e(R_{\rm{eq}} + w\xi)\cos\phi
\end{array}
\right.
\end{equation}
where $(R_{\rm{eq}}+w\xi)$ is the length of the semi-minor axis, which varies from $R_{\rm{eq}}-3w$ to $R_{\rm{eq}}+3w$ depending on the value of the model parameter $\xi$. 
The general Cartesian form of an ellipsoidal shell is derived from Eq.~\ref{eq:app2.basic_model}: 
\begin{equation}\label{eq:app2.cartesian_representation}
\dfrac{x^{2}}{(R_{\rm{eq}}+w\xi)^{2}}+\dfrac{y^{2}}{(R_{\rm{eq}}+w\xi)^{2}}+ \dfrac{z^{2}}{e^{2}(R_{\rm{eq}}+w\xi)^{2}} = 1.
\end{equation}
The infinitely thin prolate ellipsoidal surface of Appendix~\ref{app1} would be recovered if the width parameter $w$ value were set to nil.

\subsubsection{Tilted model}

The inclination of the ellipsoidal shell with the line of sight and North directions require adding two additional free parameters, namely: 
(3) the inclination of the major-axis with respect to the line of sight, $i\in[0,\pi/2]$, and 
(4) the position angle, PA$\in[-\pi,\pi)$, which is the rotation angle with respect to the $x$ axis. 
In order to correctly take into account the effects of these two orientation changes, a rotation along the $x$ axis by $i$ and a rotation along the $z$ axis by PA needs to be applied. 
This is carried out by the following transformation matrix
\begin{equation}\label{eq:app2.rotation_matrix}
\mathbf{R_{T}} = \left[
\begin{array}{ccc}
\cos(PA) & \sin(PA)\cos i & \sin(PA)\sin i\\
-\sin(PA) & \cos(PA)\cos i & \cos(PA)\sin i\\
0 & -\sin i & \cos i
\end{array}
\right]
\end{equation}
So given the position vector $\vec{P}=[x,y,z]^{\rm{T}}$ of a point within the spheroidal shell we can obtain the position vector in the observer's reference system $\vec{P'} = \mathbf{R_{T}}\vec{P}$ with $z'$ going along the line of sight.

\subsubsection{Line-of-sight velocity}

The information along the $z$ axis (line of sight) derived from the model described above can be compared with the velocity along the line of sight (or simply the radial velocity $V_r$) and velocity dispersion maps in Fig.~\ref{fig:2Dfit}.  
As described in Appendix~\ref{app1}, there are relationships between $V_{\rm eq}$ and $V_{\rm p}$ with the distance, ellipticity, angular size along the minor axis, and age of the shell (Eqs.~\ref{eq.app.veq} and \ref{eq.app.vp}).

Given a set of free parameters $\{R_{\rm{eq}}, e, i, \mathrm{PA}\}$ defining the shell geometry and space orientation, the velocity distribution of an homologously expanding shell can be modeled introducing an additional free parameter, (5) the equatorial velocity of the shell\footnote{
Contrary to Appendix~\ref{app1}, where the distance was adopted as a free parameter, the equatorial expansion velocity, $V_{\rm eq}$, is preferred here because its allows a straightforward comparison with the velocity maps. 
}, $V_{\rm{eq}}$. 
In a homologously expanding shell, the velocity at a given point has a radial direction and is proportional to the distance to the center $\vec{V}\propto \vec{P}$.  
Thus it can be proven that
\begin{equation}\label{eq:app2.velocity}
    \vec{V} = \dfrac{V_{\rm{eq}}}{R_{\rm{eq}}}\vec{P} \quad\longrightarrow\quad 
    \vec{V'} = \dfrac{V_{\rm{eq}}}{R_{\rm{eq}}}\vec{P'}=\dfrac{V_{\rm{eq}}}{R_{\rm{eq}}}\mathbf{R_{T}}\vec{P}
\end{equation}
where $V_{\rm{eq}}$ is taken in the middle of the shell ($\xi=0$). 
In this way the radial velocity $V_r$ is simply the $z$ component of the velocity in the observer's reference system
\begin{equation}\label{eq:app2.los_velocity}
    V_r = \vec{V'}\cdot \hat{k'} = \dfrac{V_{\rm{eq}}}{R_{\rm{eq}}} \left[ z\cos i - y\sin i\right]
\end{equation}
The free parameters of the model will be fitted by comparing the model 2D velocity map predictions with the observed radial velocity maps $V_r$ in Fig.~\ref{fig:2Dfit}.  
The velocity dispersion maps in Fig.~\ref{fig:2Dfit} will be used to weight the distance between the synthetic and observed velocity maps at each spaxel.

\subsubsection{Synthetic observations}

\begin{figure}[t]
\begin{center}
\includegraphics[width=1.0\linewidth]{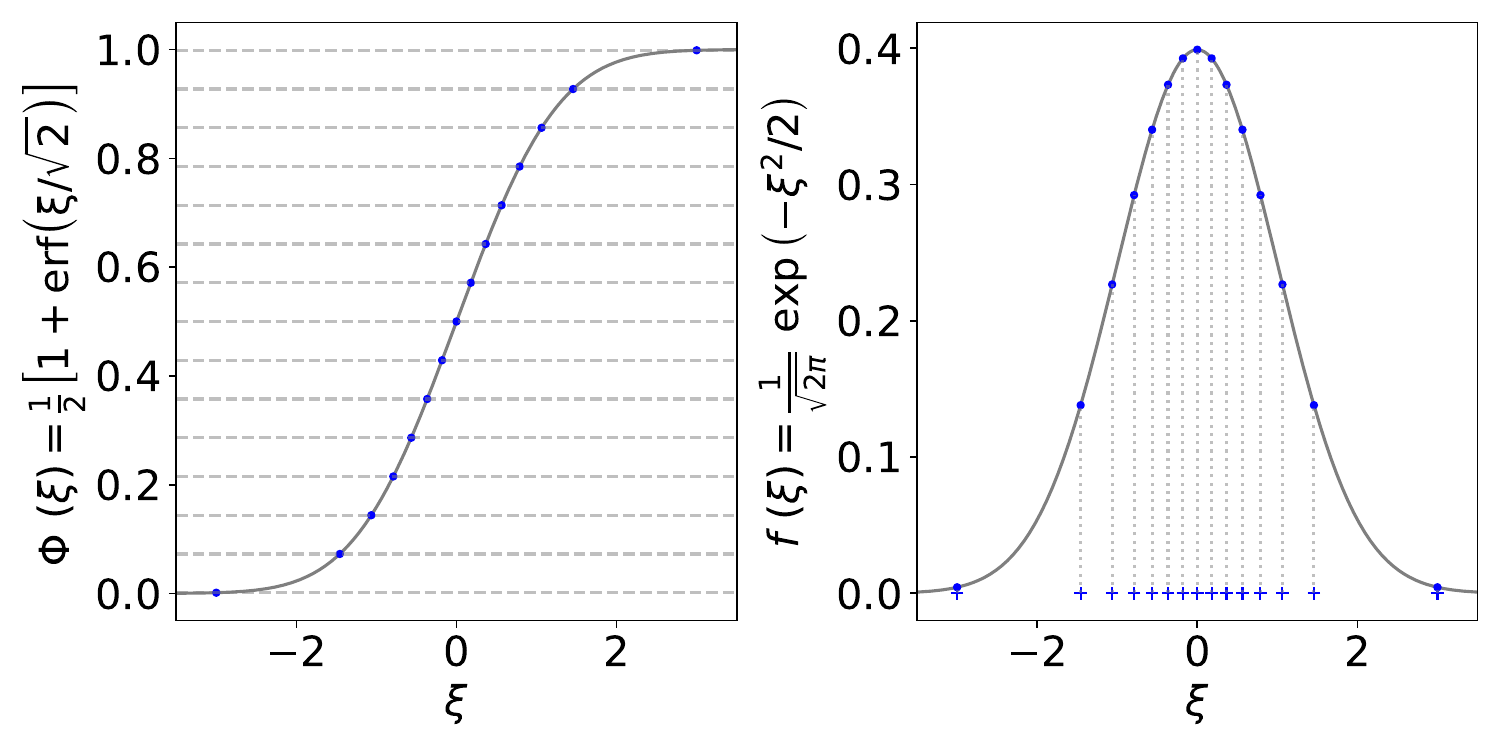}
\caption{
Distribution of the width modulator $\xi$ for 15 layers. (Left) The solid line corresponds with the normal cumulative distribution function, the dashed lines corresponds with the intervals of equal probability, meanwhile the blue dots corresponds with the intersections of the function with the intervals.
(Right) The solid line corresponds with a standard normal distribution, the blue dots are the value of the function evaluated in the intersections of the equally probable intervals, meanwhile the crossed points in the bottom shows qualitatively the separation between the layers in the model with respect to the center of the shell.
}
\label{fig:app2.xi_distribution}
\end{center}
\end{figure}

Given a set of the free parameters $\{R_{\rm{eq}}, e, V_{\rm{eq}}, i, PA\}$, the model parameters $\{\xi, \theta, \phi\}$ are used to sample points within the spheroidal shell. 
To represent significantly the model, the layering approach will use a width modulator $\xi$ discretized by means of the normal cumulative distribution function defined as 

\begin{equation}\label{eq:app2.normal_cumulative_function}
    \Phi\left(x\right) = \dfrac{1}{\sqrt{2\pi}}\int_{-\infty}^{x}e^{-\frac{t^{2}}{2}}~dt = \dfrac{1}{2}\left[1 + \erf\left(\dfrac{x}{\sqrt{2}}\right)\right]
\end{equation}

\noindent being $\erf(z)$ the standard error function 

\begin{equation}\label{eq:app2.error_function}
    \erf\left(z\right) = \dfrac{2}{\sqrt{\pi}}\int_{0}^{z}e^{-t^{2}}~dt
\end{equation}

\noindent in order to $\xi$ being a normally distributed value as seen in the right panel of Fig.~\ref{fig:app2.xi_distribution}.  
This is achieved by uniformly splitting the range of the cumulative distribution function (Eq.~\ref{eq:app2.normal_cumulative_function}) into $(n_{\rm{layers}}-1)$ intervals as shown in the left panel of the Fig.~\ref{fig:app2.xi_distribution} and finding the value of $\xi$ in every intersection. Here $n_{\rm{layers}}$ is the number of modeled layers.

Thus for every value of $\xi_i$, there is an ellipsoidal surface of semi-minor axis $b_i=R_{\rm{eq}}+w\xi_i$. 
Since every surface has different area, in order to conserve a point surface density, every layer needs to have a number of points given by relationship 
\begin{equation}\label{eq:app2.points_per_layer}
 n_{i} = \left\lfloor n_{1}\left(\dfrac{R_{\rm{eq}}+w\xi_{i}}{R_{\rm{eq}}+w\xi_{1}}\right)^{2}\right\rfloor
\end{equation}
where $\lfloor\cdot\rfloor$ represents the standard floor operation, which rounds down the value inside the brackets to the nearest integer, and $2n_{i}+1$ is the number of points in the $i$-th layer.  
The minimum number of points per layer is $2n_{1}+1$ for a value of $i=1$.

Within each layer, the points where uniformly distributed in the surface by means of a Fibonacci lattice, which has a better spatial covering properties than the longitude-latitude lattice \citep{Gonzalez2010}. 
To sample the polar and azimuthal angles a variation of the equations shown in \citet{Gonzalez2010} was used 
\begin{equation}\label{eq:app2.fibonacci_lattice_angles}
\begin{array}{l}
    \theta_{i,j} = 2\pi \left({\dfrac{j}{\varphi} \mod 1}\right) + 2\pi\dfrac{(i-1)}{(n_{\rm{layers}}-1)} \\~\\
    \phi_{i,j}  = \dfrac{\pi}{2} - \arcsin\left(\dfrac{2j}{2n_{i}+1}\right) + \pi\dfrac{(i-1)}{(n_{\rm{layers}}-1)}\\~\\
    j\in\{-n_{i}, \dots,0\dots, n_{i}\}
\end{array}
\end{equation}
where $\varphi=\left(1+\sqrt{5}\right)/2$ is the golden ratio. 
This variation gives for every layer a little offset to both angles in order to prevent the polar under density of points and improve the spatial sampling. 
In this work we used $n_{\rm{layers}}=512$ and $n_{1}=2048$ for the fitting.

\begin{figure}[t]
\begin{center}
\includegraphics[width=\linewidth]{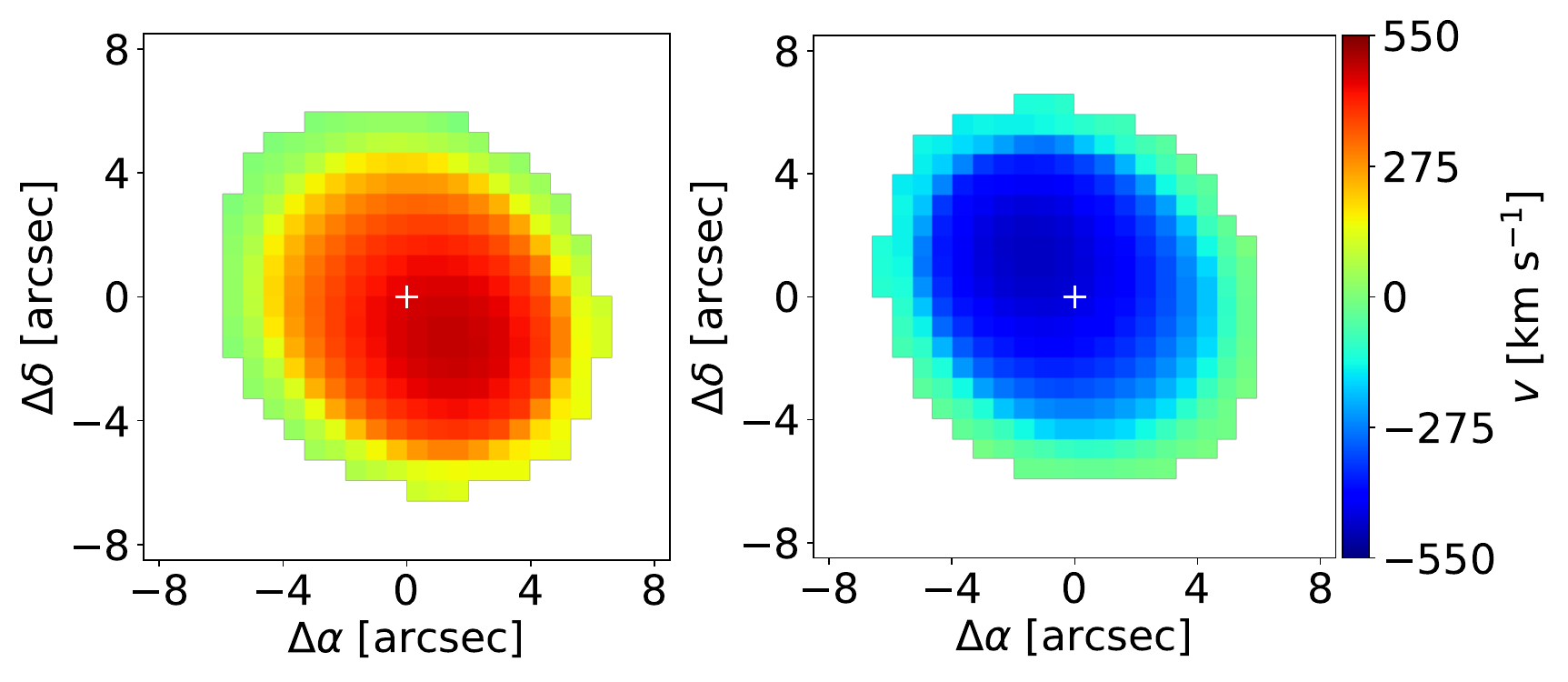}
\caption{
Example of a synthetic observation produced by the model with the set of parameters $R_{\rm{eq}}=5\farcs0$, $w=0\farcs13$, $e=1.4$, $V_{\rm{eq}}=400~\rm{km~s^{-1}}$, $i=45^{\circ}$, $PA=45^{\circ}$. 
The field of view and spatial resolution are the same as in Fig~\ref{fig:2Dfit}. 
The center is marked by a white cross.
}
\label{fig:app2.example_prediction_of_model}
\end{center}
\end{figure}

Once the model is sampled, 2D histograms of the mean velocity per bin were made as shown in Figure~\ref{fig:app2.example_prediction_of_model}. 
Those histograms have the same pixel size and spatial resolution as the VLT VIMOS observations to allow a fair comparison of the model with the real data.

\subsubsection{Observed data filtering}

\begin{figure}[t]
    \begin{center}
\includegraphics[bb=15 40 725 1350,width=0.90\linewidth]{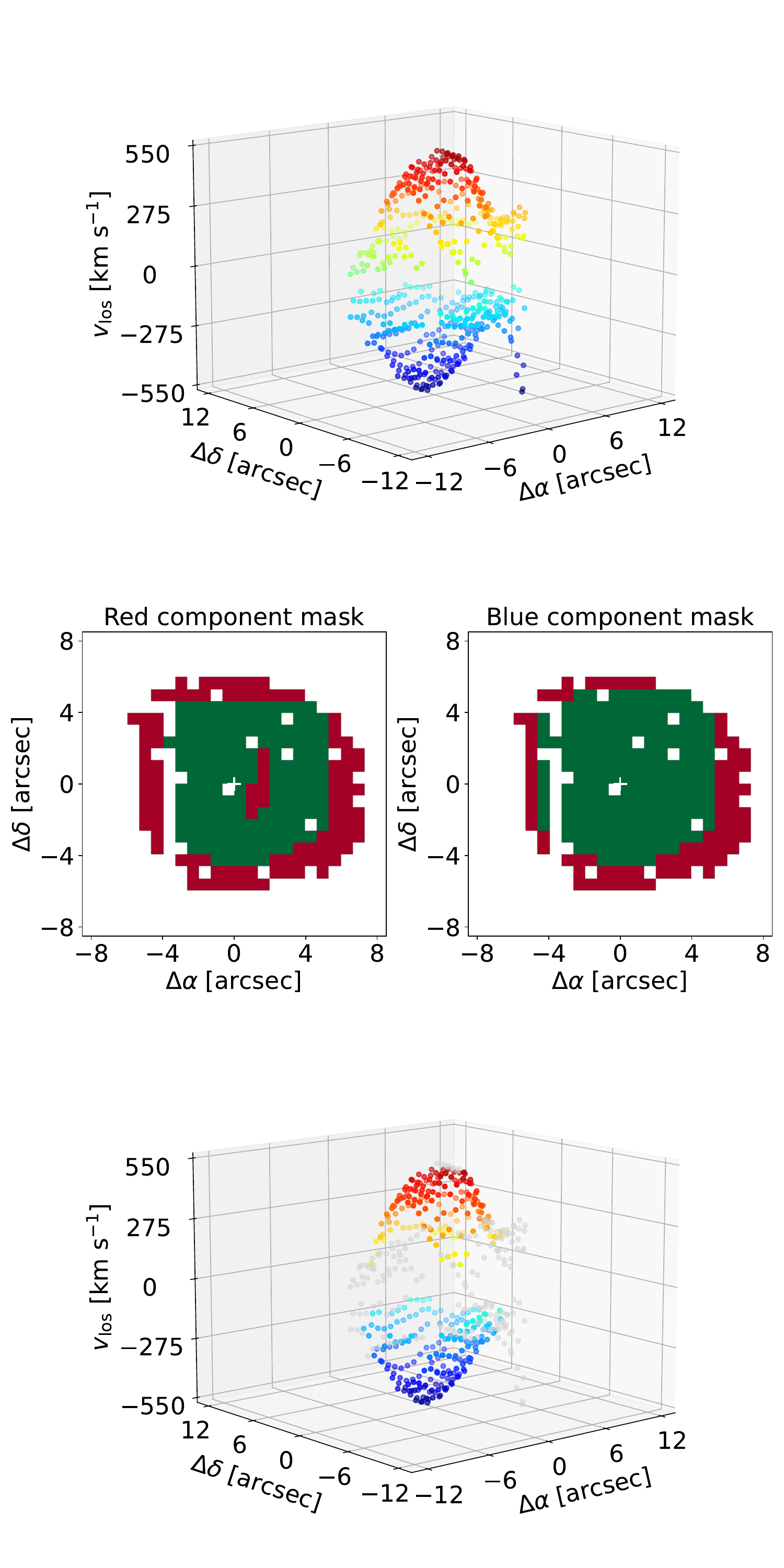}
\caption{
(top) 
3D scatter plot of the velocity observations. 
The spaxels along the edge of the shell shows noticeable excursions in velocity, as well as a "swimming ring-like" feature. 
(middle) 
Pixels used in the MCMC procedure (green) and those masked out (red) because the derived value of $V_{\rm{los}}$ differs notably for an expanding spheroidal shell. 
The center used is located at position (0,0) as marked with a white cross.
(bottom)
3D scatter plot of the velocity observations once the mask is applied (grey dots). 
The remaining data points resemble a spheroidal-like structure.
}
\label{fig:app2.pixel_mask}
\end{center}
\end{figure}

The 3D visualization of the maps of the red and blue velocity components at the top-panel of Fig.~\ref{fig:app2.pixel_mask} reveals a number of spaxels along the outer edge whose behavior differ notably from that expected for an expanding ellipsoidal shell.  
Velocity excursions and a ``kind of swimming ring'' at the ellipsoid equator are obvious in this figure.  
This effect is caused by the problematic two Gaussian fit to the velocity profile at the edge of the shell, where the red and blue velocity components blend together. 
In addition, the red component showed a hump-like feature at its maximum radial velocity.  
This is actually caused by contamination from the [N~{\sc ii}] ring-like structure (see the middle and bottom panels of Fig.~\ref{fig:3d}).  

The velocity dispersion maps were used to downplay these effects by giving a lower weight in the fit to spaxels with larger velocity dispersion, but the effects of these spaxels on the fit were still noticeable.  
In order to definitely alleviate the impact of these spaxels, they were disregarded in the fit using the mask shown in the middle-panel of Fig.~\ref{fig:app2.pixel_mask}. 
Similarly, since the velocity information of the spaxels affected by the contamination from the [N~{\sc ii}] ring-like structure do not fully represent that of the expanding ellipsoid, these spaxels have also been disregarded for the fit.  
The 3D visualization of the useful maps of the red and blue velocity components is shown at the bottom-panel of Fig.~\ref{fig:app2.pixel_mask}.

\subsection{Markov Chain Montecarlo fit}

In order to fit a homologous expanding prolate ellipsoidal shell  model, a MCMC (Markov Chain Montecarlo) approach was implemented using the \texttt{Python} library \texttt{emcee} \citep{emcee2013}.
The details of this procedure are explained below.

\begin{figure}[t]
    \begin{center}
\includegraphics[width=\linewidth]{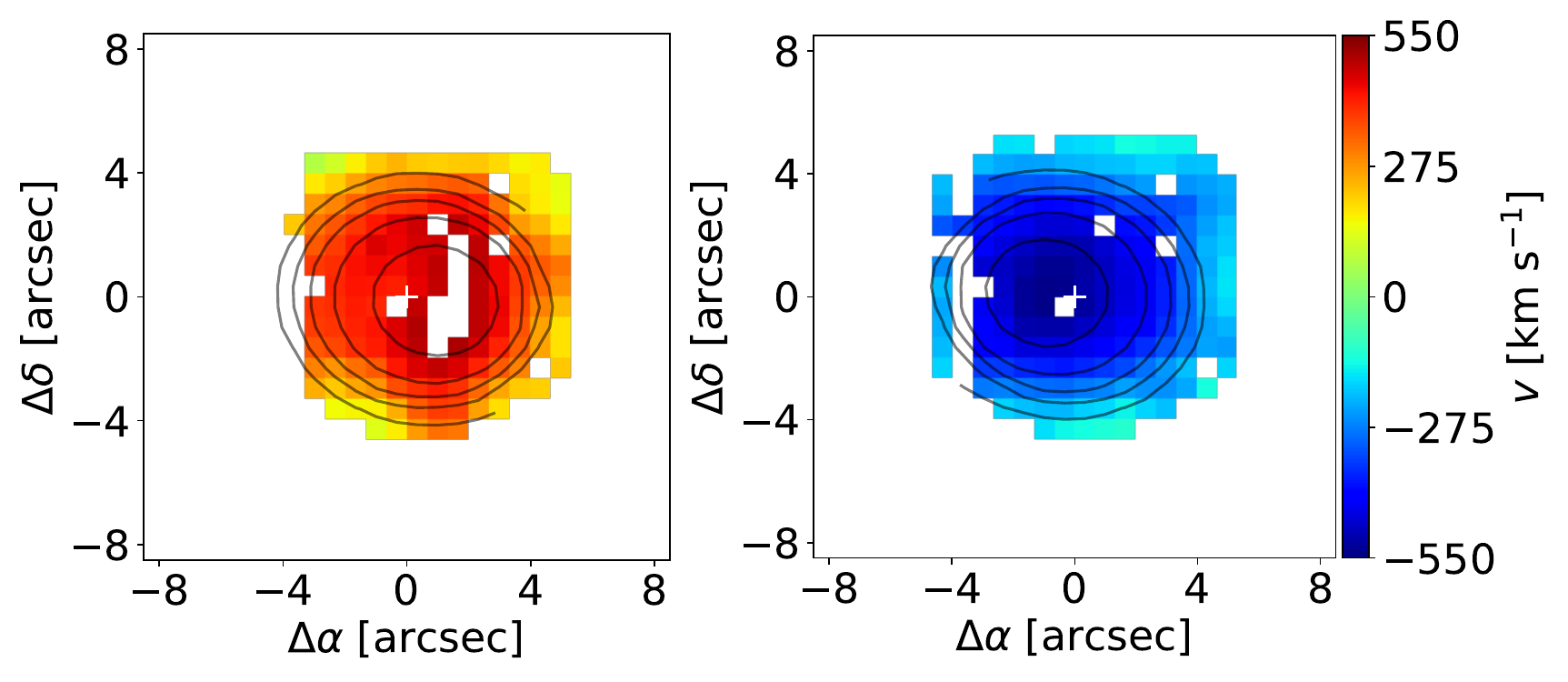}
\includegraphics[width=\linewidth]{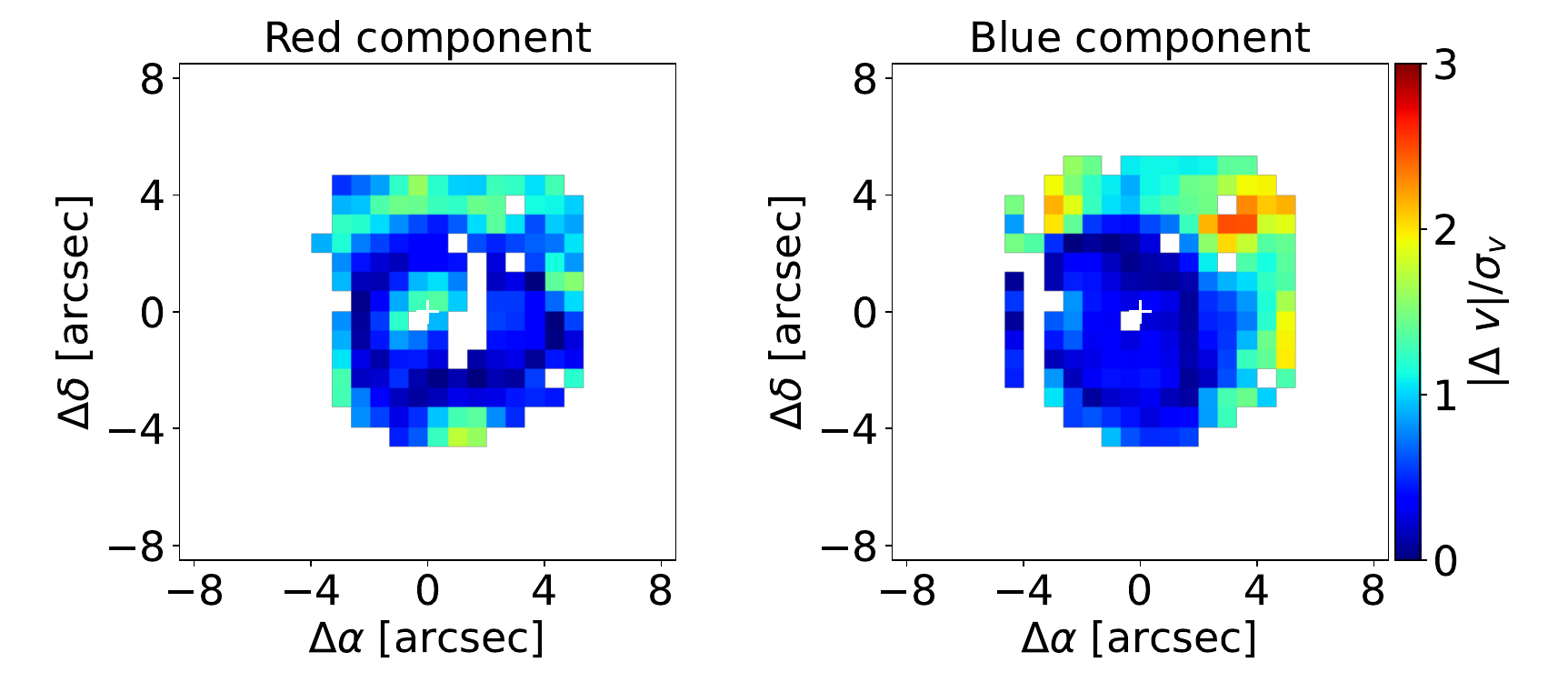}
\caption{
(top) MCMC best-fit model for the red (left) and blue (right) velocity components of the expanding ellipsoid shown as contours overimposed on the observed velocity maps shown in Fig.~\ref{fig:2Dfit}, and (bottom) absolute differences between the MCMC best-fit model and the velocity maps in Fig.~\ref{fig:2Dfit} with respect to each image pixel velocity dispersion. 
The center is located at position (0,0) as marked with a white cross. 
}
\label{fig:app2.fitted_model}
\end{center}
\end{figure}

\subsubsection{Likelihood function}

The comparison of the synthetic blue and red components of the radial expansion velocity with the observed ones in Fig.~\ref{fig:2Dfit} was made on a pixel-by-pixel basis. 
In this way the likelihood function is defined as
\begin{equation}
\ln\,p(v\,|\,R_{\rm{eq}},w,\epsilon,V_{\rm{eq}},i,PA) =
    -\frac{1}{2} \sum_i \left[
        \frac{(v_i-v_{i}')^2}{s_i^2}
        + \ln \left ( 2\pi\,s_i^2 \right )
    \right], 
\end{equation}
where $i$ represents a pixel and $v_{i}$, $s_{i}$ and $v_{i}'$ correspond to the observed velocity, observed velocity dispersion, and model predicted velocity in pixel $i$, respectively. 
The addition on the right term of this equation over all pixels is made considering simultaneously the red and blue velocity components. 
It must be remarked that only valid pixels, as masked according to Fig.~\ref{fig:app2.pixel_mask}, are considered here.

\subsubsection{Parameter's space}

The prior distribution for every parameter is adopted to be flat (uninformed prior distribution), since there is no initial information about them. 
Otherwise every parameter has its respective domain, but a feasible fit of the observations via MCMC requires making a good guess of them. 
The observed distribution's edge of the shell seems to be around 5~arcsec (Fig,~\ref{fig:2Dfit}), thus it has been constrained to be $R_{\rm{eq}}\in[4'',6'']$.  
Meanwhile the shell width $w$ was fixed at a value of 0.13 arcsec, as described above. 
The axial ratio $e$ was constrained to be greater than 1; $e\in[1,\infty)$, whereas for the inclination and position angles, all their physical range were taken into account, i.e., $i\in[0,\pi/2]$ and $PA\in[-\pi, \pi)$. 
Finally the equatorial velocity can be expected to be in the range $V_{\rm{eq}}=300-600~\rm{km~s^{-1}}$.

In order to fit the model to the observations a set of 128 random walkers were initialized based on a uniform probability distribution of the parameter's ranges described before.
We note that an additional constraint was applied to the possible values of $e$, $i$, and $R_{\rm eq}$, so that the projected axial ratio was in the range from 1.1 to 1.2, in agreement with the observed axial ratio of the nova remnant measured in the HST image.

\begin{figure}
\begin{center}
\includegraphics[width=1.00\linewidth]{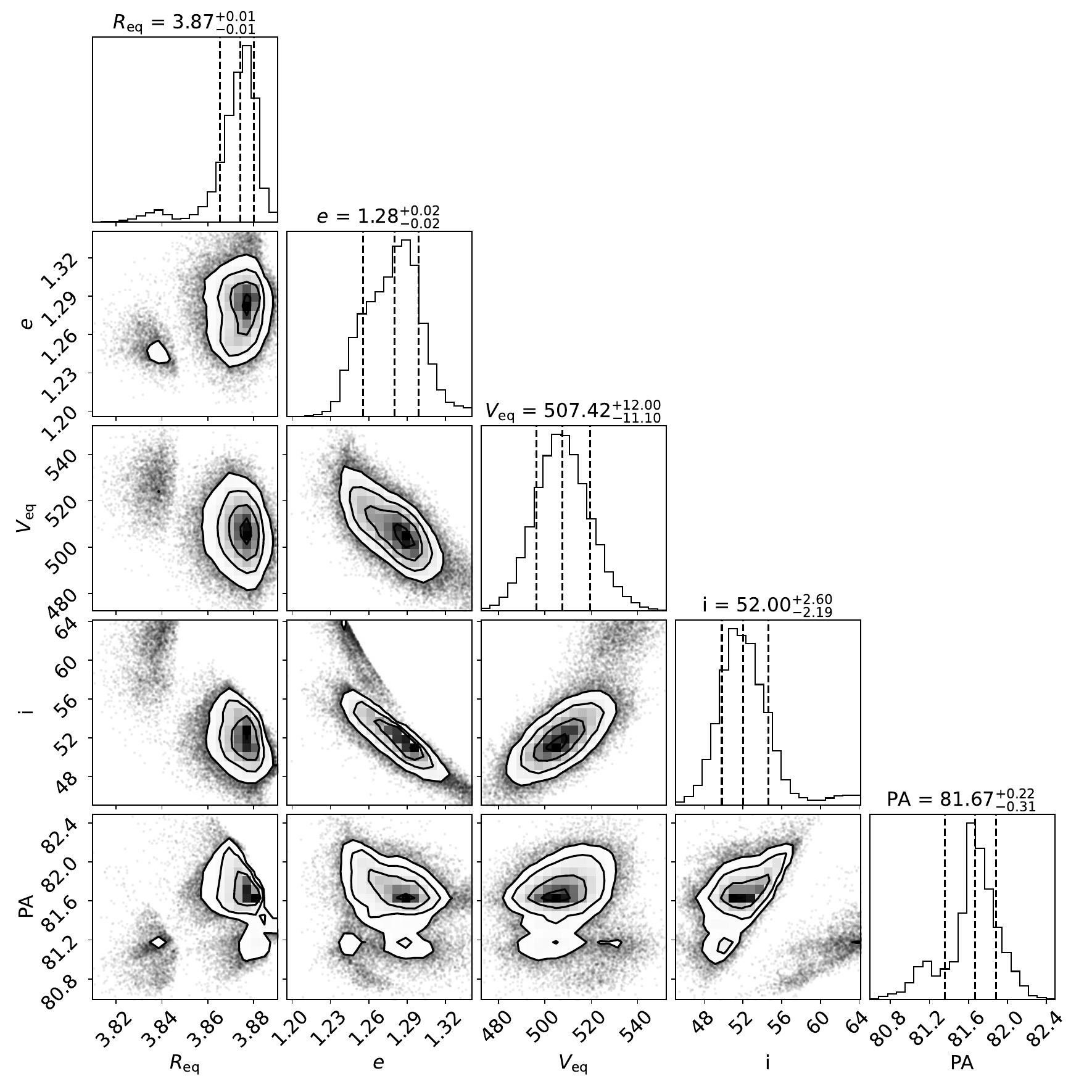}
\caption{
Corner plot for the resulting parameters of the MCMC approach fitted to the observations. 
Only the parameter's space corresponding to $3\sigma$ is used in the histograms.
}
\label{fig:app2.corner_plot}
\end{center}
\end{figure}

\subsection{Best-fit model}

The MCMC best-fit model and the associated difference between the red and blue velocity components are shown in Fig.~\ref{fig:app2.fitted_model}.  
The bottom panels of that figure clearly show that the differences between the expected model velocities at each pixel and those actually observed differ by less than the velocity dispersion for most pixels.

The corner plot with the distribution of the best-fit parameters is shown in Fig.~\ref{fig:app2.corner_plot}.  
Their best values and 1-$\sigma$ uncertainties are labeled on top of each row.  
Some parameters, namely $e$, $V_{\rm eq}$ and $i$, exhibit some correlation, which can be expected.  
The other parameter combinations seem rather uncorrelated.

The best-fit parameters imply equatorial radius and velocity of 3.87$\pm$0.01 arcsec and 507$\pm$12 km~s$^{-1}$, respectively.  
The axial ratio is found to be 1.28$\pm$0.02, whereas the inclination is 52$^\circ\pm$2$^\circ$.  
The polar velocity would then be 649$\pm$26 km~s$^{-1}$. 
These best-fit parameters are within 1-$\sigma$ of those derived in Appendix~\ref{app1}, but for the axial ratio, which is a bit larger, 1.28$\pm$0.02 against 1.22$\pm$0.03, but still marginally consistent. 
Finally it must be noted that the best-fit value of the PA, 81.7$^\circ\pm$0.2$^\circ$, differs with respect to the PA of the shell of 86$^\circ$ described in Section~2.3.1, which reflects the limited spatial information of the velocity maps used to constrain the fit.

\begin{figure}[h]
\begin{center}
\includegraphics[width=0.90\linewidth]{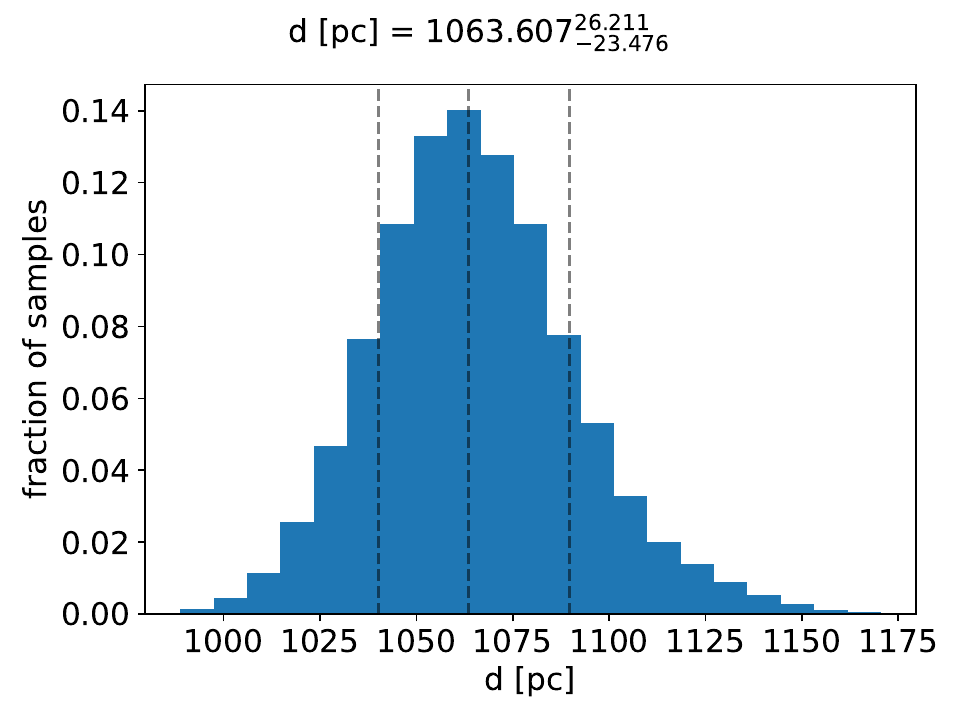}
\caption{
Distribution of the derived distances using the corresponding velocity distribution from MCMC. Calculated using $R_{\rm{eq}}$ and $V_{\rm{eq}}$ from the MCMC fit.
}
\label{fig:app2.distance_distribution}
\end{center}
\end{figure}

The information on $R_{\rm eq}$ and $V_{\rm eq}$ provides a distribution for the distance (Fig.~\ref{fig:app2.distance_distribution}), which is found to be 1064$\pm$26 pc, in agreement with the Gaia estimate of 1080$^{+70}_{-90}$ pc.

\label{lastpage}
\end{document}